\documentclass[prb,twocolumn,superscriptaddress,citeautoscript,showpacs,amsart,longbibliography]{revtex4}

\usepackage{graphicx}
\usepackage{multirow}
\usepackage{color}
\usepackage{bm}
\usepackage{times}
\usepackage{amsmath,bm,amsfonts}
\usepackage{dcolumn}
\usepackage{graphicx}
\usepackage{latexsym}

\usepackage{ulem} 
\usepackage{braket}
\usepackage{mathtools}

\usepackage{cancel}

\begin{document}


\title{Topological Magnetoelastic Excitations in Non-Collinear Antiferromagnets}

\author{Sungjoon \surname{Park}}

 \affiliation{Department of Physics and Astronomy, Seoul National University, Seoul 08826, Korea}

\affiliation{Center for Correlated Electron Systems, Institute for Basic Science (IBS), Seoul 08826, Korea}

\affiliation{Center for Theoretical Physics (CTP), Seoul National University, Seoul 08826, Korea}

\author{Bohm-Jung \surname{Yang}}
\email{bjyang@snu.ac.kr}
 \affiliation{Department of Physics and Astronomy, Seoul National University, Seoul 08826, Korea}

\affiliation{Center for Correlated Electron Systems, Institute for Basic Science (IBS), Seoul 08826, Korea}

\affiliation{Center for Theoretical Physics (CTP), Seoul National University, Seoul 08826, Korea}

\date{\today}

\begin{abstract}We study the topological property of the magnetoelastic excitation in non-collinear antiferromagnets. As a toy model, we consider the magnon-phonon coupling in a triangular antiferromagnet with a $120^\circ$ N\`eel order. 
We find that in the presence of out-of-plane external magnetic field, the magnon-polaron bands, which arise from hybridization of magnons and phonons, can carry Chern number, even though the individual magnon and phonon bands are topologically trivial. Large Berry curvature is induced from the anti-crossing regions between the magnon and phonon bands, which renormalizes the thermal Hall conductivity of phonon bands. To compute the Berry curvature and Chern number of magnon-polarons, we give a simple algorithm to diagonalize magnetoelastic Hamiltonian without diagonalizing the phonon Hamiltonian, by mapping the problem to the diagonalization of bosonic Bogoliubov-de-Gennes (BdG) Hamiltonian. This is necessary because the contribution to the Berry curvature from phonon cannot be properly captured if we compute the Berry curvature from magnetoelastic Hamiltonian whose phonon sector has been already diagonalized.
\end{abstract}

\maketitle

\section{Introduction}

Since the discovery of the quantum Hall effect \cite{klitzing1980new,laughlin1981quantized,haldane1988model}, the role of topology in electronic systems has been extensively researched. 
Recently, the implications of nontrivial topology has also been investigated in bosonic quasiparticles such as magnons\cite{shindou2013topological} and phonons \cite{zhang2010topological,zhang2011phonon}. 
It was found that although the Chern number does not guarantee a quantized response as in fermions because of the nature of bosonic statistics, non-zero Chern number still indicates the presence of chiral edge modes \cite{shindou2013topological} and the non-zero Berry curvature contributes to magnon\cite{katsura2010theory,onose2010observation,matsumoto2011theoretical,matsumoto2011rotational} and phonon \cite{strohm2005phenomenological,sheng2006theory,inyushkin2007phonon,zhang2010topological,zhang2011phonon,qin2012berry} thermal Hall effect,  and magnon spin Nernst effect\cite{kovalev2016spin,zyuzin2016magnon}

On the other hand, it has long been known that magnon can couple naturally to phonons in ferromagnets and antiferromagnets \cite{kittel1949physical,kittel1958interaction,callen1963static} . The source of this coupling can be roughly put into two categories\cite{jones1966magnetoelastic}, the first of which arises from inter-ionic  spin-spin interactions, such as strain-variation of dipole-dipole interactions and exchange interactions. The second category arises from intra-ionic spin-orbit interaction, wherein the spins sense the variation of crystal field that arise from strain via spin-orbit coupling. Regardless of the origin, when magnetoelastic coupling term that is quadratic in magnon and phonon operators does not vanish, magnon and phonon can hybridize to form a quasi-particle that is an admixture thereof,  \cite{kittel1958interaction} which has been termed `magnon-polaron' \cite{shen2015laser,kamra2015coherent}. 

Recently, various phenomena rooted in magnetoelastic coupling in ferromagnets have been studied, with potential applications in spin and phonon control. In Refs.~[\onlinecite{uchida2011long,weiler2012spin,kamra2015coherent,xu2018inverse}], it was proposed that the magnon-phonon coupling in ferromagnets can be utilized in spintronics by exploiting acoustic spin pumping. In Ref.~[\onlinecite{nomura2018phonon}], it was shown that phonon velocity propagating parallel to and antiparallel to external magnetic field can differ due to magnetoelastic coupling, which may find usage in phononics. In Ref.~[\onlinecite{takahashi2016berry}], it was proposed that large Berry curvature can be induced in the anti-crossing regions of magnon and phonon bands, which can be utilized to control magnon current. It was also suggested that magnon-phonon coupling contributes significantly to Hall conductivity in response to gradient in external magnetic field \cite{thingstad2018chiral} as well as spin and thermal conductivities \cite{flebus2017magnon}. 

In contrast, magnetoelastic coupling in antiferromagnets has been relatively less studied. However, recent experiments showed that magnon-phonon coupling can be large in hexagonal rare-earth manganite $\textrm{RMnO}_3$ ($\textrm{R=Y,}~\textrm{Lu},~\textrm{Ho}$), which are approximately triangular antiferromagnets\cite{oh2016spontaneous,kim2018renormalization}. In Ref.~[\onlinecite{oh2016spontaneous}], the authors showed that magnetoelastic coupling contribute significantly to magnon decay for R=Y, Lu, and in Ref.~[\onlinecite{kim2018renormalization}], the authors showed that magnetoelastic coupling can significantly renormalize magnon spectrum for R=Ho. 
Since the magnetoelastic coupling accompanies the anticrossing between magnon and phonon bands, one can expect novel topological phenomena to arise in hybridized band structure.

In this paper, we examine the topological property of magnon-polaron bands in a triangular antiferromagnet with a $120^\circ$ N\`eel order. 
Although the ground state configuration enlarges the unit cell, the magnetic excitation keeps the translation symmetry of the underlying triangular lattice.  
Thus, there is only one magnon band in the Brillouin zone, and one cannot expect any topological property in the magnon band.
However, once the magnon-phonon coupling is considered, the hybridized band structure with three magnetoelastic bands can support non-trivial band topology. 
We find that the magnetoelastic coupling arising from exchange striction does not open all of the gaps between the magnon and phonon. 
However, the application of external magnetic field removes all of the gap closing points, resulting in topological magnon-polaron bands with non-zero Chern number.

In addition, in order to calculate the Berry curvature and the Chern number of magnon-polaron bands, we develop a method to diagonalize the magnetoelastic Hamiltonian.  
This step is necessary because, although the magnetoelastic Hamiltonian is often written in the Holstein-Primakoff (HP) operator and phonon operator basis\cite{takahashi2016berry,flebus2017magnon,thingstad2018chiral}, calculating the Berry curvature in this basis does not give the correct Berry curvature for the magnon-polarons. 
The reason is that if we write the magnetoelastic Hamiltonian using phonon basis, the phonon Hamiltonian is already diagonalized, so that the Berry curvature computed in this way cannot correctly capture the contribution from the phonon wave function.
We find that the problem of diagonalizing magnetoelastic Hamiltonian can easily be solved by observing that the phonon Hamiltonian can be mapped to a bosonic BdG Hamiltonian by a simple transformation of basis. 
Thus, if we also write the magnon Hamiltonian in BdG form, the magnetoelastic Hamiltonian is also in BdG form, for which the problem of diagonalizing the Hamiltonian and computing the Berry curvature is well known \cite{colpa1978diagonalization,shindou2013topological}.

This paper is organized as follows. In Sec.~\ref{sec.mptriangular}, we study the energy spectrum of magnon-polaron on triangular lattice. We show that in the presence of external magnetic field, all of the magnon-polaron bands become decoupled. In Sec.~\ref{sec.bc_tc}, we compute the Berry curvature and the thermal hall conductivity, and show that the decoupled bands carry non-zero Chern numbers. In Sec.~\ref{sec.diagonalization}, we present a general formalism to diagonalize the magnetoelastic Hamiltonian, which is written using HP operators in magnon sector, and displacement and momentum operators in the phonon sector. This method should be compared with the method where magnetoelastic Hamiltonian is written with the phonon operators. Although the two methods give the same energy spectrum, their Berry curvatures are different, as explained in Sec.~\ref{sec.berry}. We conclude in Sec.~\ref{sec.conclusion}.

\section{Magnon-polaron spectrum in Triangular Antiferromagnet} \label{sec.mptriangular}

In this section, we present a toy model of topological magnon-polaron in a triangular antiferromagnet.
We begin by examining the magnon spectrum and symmetries of Heisenberg triangular antiferromagnet, and then introduce easy-axis anisotropy and external magnetic field. Then, we study the phonon spectrum in triangular lattice with external magnetic field. Finally, we turn on the interaction between magnons and the in-plane vibrations, which can naturally arise in non-collinear antiferromagnets, as will be explained below. In the presence of magnetic field and the magnon-phonon coupling, all of the bands decouple from each other.

\begin{figure}[ht]
\centering
\includegraphics[width=8cm]{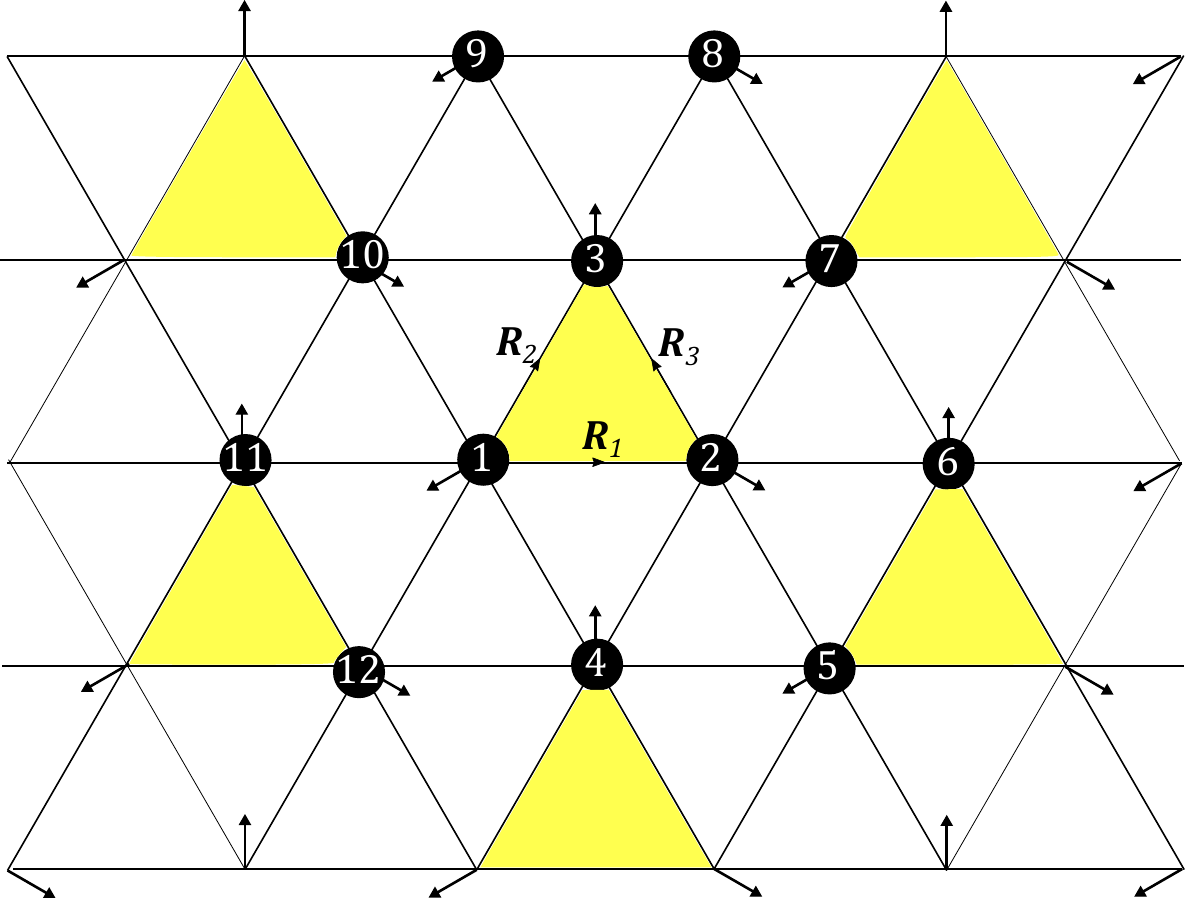}
\caption{Triangular Heisenberg antiferromagnet with $120^\circ$ N\`eel order, in which the spins rotate by $120^\circ$ counterclockwise for translations by $\bm{R}_1$, $\bm{R}_3$, and $-\bm{R}_2$. Because of the magnetic ordering, the unit cell of the magnetic ground state is enlarged, as indicated by the yellow triangles. However, the translation symmetry of the triangular lattice without magnetic order is restored in the magnon spectrum. We have labeled some of the lattice sites for convenience.}
\label{fig.triangular_lattice}
\end{figure}

\subsection{Magnon} \label{ssec.mtriangular}

Let us study the magnon spectrum on a triangular lattice with the Hamiltonian given by
\begin{equation}
{\cal H}_m={\cal H}_J+{\cal H}_A+{\cal H}_H,
\end{equation}
where ${\cal H}_J$ is the antiferromagnetic Heisenberg interaction, ${\cal H}_A$ is the easy axis anisotropy, and ${\cal H}_H$ is the coupling  to the external magnetic field. Below, we will study each term separately. The antiferromagnetic Heisenberg Hamiltonian is given by 
\begin{equation}
{\cal H}_{J}=J\sum_{\langle ij \rangle} \bm{S}_{i}\cdot \bm{S}_{j},
\end{equation} 
where $J>0$, and the summation is over the nearest neighboring spins.
Its ground state is the $120^\circ$ N\`eel state\cite{capriotti1999long,zheng2006excitation,white2007neel} shown in Fig.~\ref{fig.triangular_lattice}.

The magnon Hamiltonian can be found by introducing local coordinates for each of the spins and by introducing the HP operators with respect to the local coordinates. We always choose the local $z$-axis to point in the direction of the classical magnetic order. We choose the local $y$ axis to point out of the plane, which leaves only one possibility for the local $x$ axis. Then, we write $\bm{S}_i=S^x_i \hat{\bm{x}}_i+S^y_i \hat{\bm{y}}_i+S^z_i \hat{\bm{z}}_i$, where $\hat{\bm{x}}_i,\hat{\bm{y}}_i,\hat{\bm{z}}_i$ are the \textit{local} axes for the spin at position $i$.  We find
\begin{align}
\bm{S}_{i} \cdot \bm{S}_{j} =& S_{i}^y S_{j}^y+\cos(\theta_{i}-\theta_{j})(S_{i}^z S_{j}^z + S_{i}^x S_{j}^x) \nonumber \\
&+\sin(\theta_{i}-\theta_{j})(S_{i}^z S_{j}^x -S_{i}^x S_{j}^z),
\label{eq.s1dots2}\end{align}
where $\theta_i$ is measured with respect to the \textit{global} $x$-axis, which is parallel to $\bm{R}_1$ in Fig.~\ref{fig.triangular_lattice}.
The HP transformation with linear spin wave approximation is $S_i^z=S-a_i^\dagger a_i$, $S_i^x=\frac{\sqrt{2s}}{2}(a_i+a_i^\dagger)$, $S_i^y=\frac{\sqrt{2S}}{2i}(a_i-a_i^\dagger)$. Taking the Fourier transformation
\begin{equation}
a_i=\sum_{\bm{k}}{e^{i\bm{k}\cdot \bm{R}_i}a_{\bm{k}}},
\end{equation}
where $\bm{R}_i$ is the position of the $i$th atom, we obtain
\begin{equation}
{\cal H}_J=\sum_{\bm{k}}{\left[A_{\bm{k}} a^\dagger_{\bm{k}} a_{\bm{k}} - \frac{1}{2}B_{\bm{k}}(a_{\bm{k}}^\dagger a_{-\bm{k}}^\dagger + a_{-\bm{k}}a_{\bm{k}})      \right ]},
\label{eq.heisenberg}\end{equation}
where we kept only the terms quadratic in the HP operators. Here, $A_{\bm{k}}=3JS(1+\frac{1}{2}\gamma_{\bm{k}}) \quad B_{\bm{k}}=\frac{9}{2}JS\gamma_{\bm{k}}$
and $\gamma_{\bm{k}}=\frac{1}{6}\sum_{\bm{\delta}}{e^{i\bm{k}\cdot \bm{\delta}}}$ where $\bm{\delta}$ are the vectors pointing towards the six nearest neighbors from a given site. 

\begin{figure}[t]
\centering
\includegraphics[width=8.5cm]{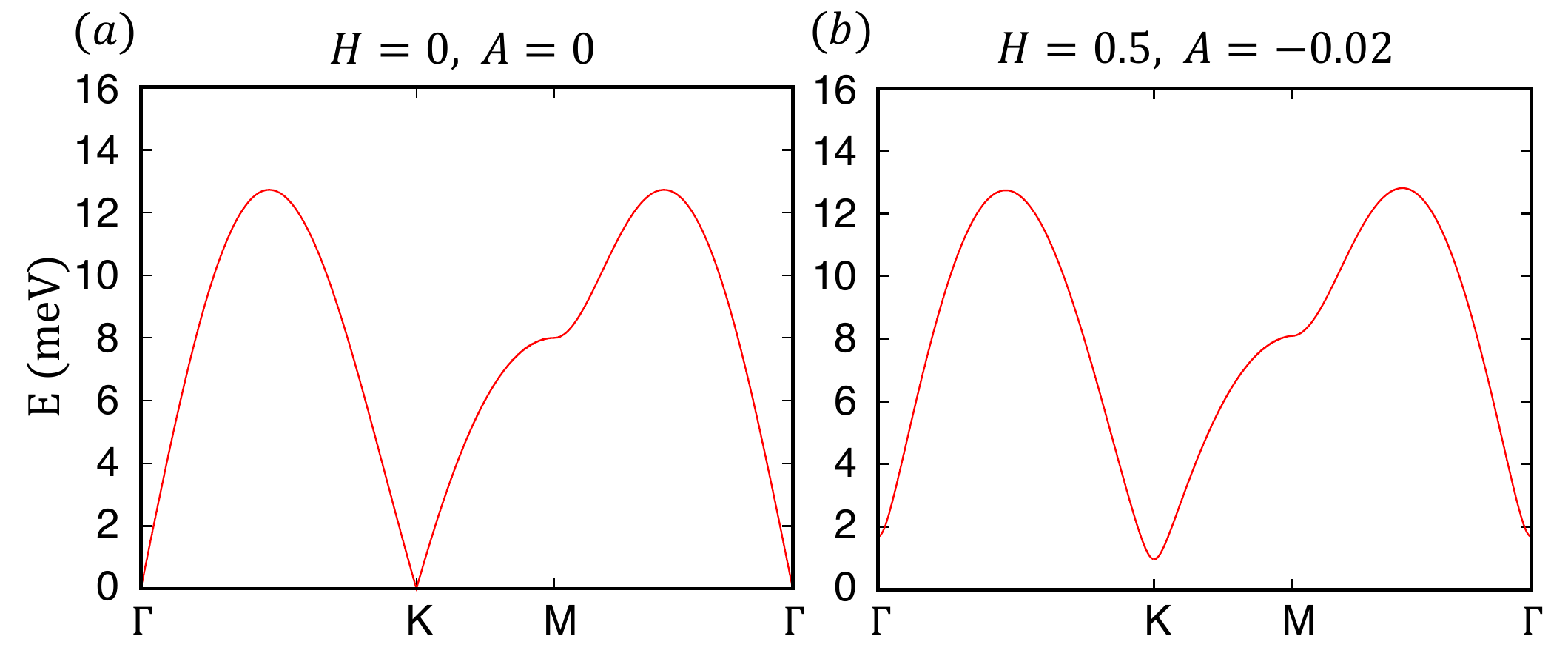}
\caption{Magnon spectrum along the high symmetry line in the unit of meV. (a) The magnon band with only the Heisenberg interaction with $J=2$ meV, $S=2$. (b) The magnon band with anisotropy $A=-0.02$ meV and magnetic field $H= 0.5$ meV. The  role of the anisotropy is to remove all of the Goldstone modes. The role of the magnetic field is to remove all of the band degeneracies between magnon and phonons.}
\label{fig.magnon_hsl}
\end{figure}

Let us note that if we define 
\begin{equation}
\phi_{\bm{k}}=\begin{pmatrix} a_{\bm{k}} \\ a_{-\bm{k}}^\dagger \end{pmatrix},
\end{equation}
and define $\tau_i$ to be the $2\times 2$ Pauli matrices that relate particle and hole,  we have the following relations, which define bosonic BdG field operators:
\begin{equation}
[\phi_{\bm{k},i}, \phi_{\bm{k},j}^\dagger ]=(\tau_z)_{ij}, \quad \phi_{-\bm{k}}=\tau_{x} \phi_{\bm{k}}^\dagger. \label{eq.heisenberg_field}
\end{equation} 
Since we can write
\begin{align}
{\cal H}_{J}&=\frac{1}{2}\sum_{\bm{k}}
\phi_{\bm{k}}^\dagger 
\begin{pmatrix}
A_{\bm{k}} & B_{\bm{k}} \\
B_{\bm{k}} & A_{\bm{k}}
\end{pmatrix} 
\phi_{\bm{k}}\nonumber \\
&=\sum_{\bm{k}}
\phi_{\bm{k}}^\dagger 
H_J(\bm{k})
\phi_{\bm{k}}, 
\label{eq.simple_magnon_example}\end{align}
where $H_J(\bm{k})$ is a bosonic BdG Hamiltonian. For notational simplicity,  we will write $\phi_{\bm{k}}^\dagger$ for either $\begin{pmatrix}
a_{\bm{k}}^\dagger \\ a_{-\bm{k}}
\end{pmatrix}$ or $\begin{pmatrix}
a_{\bm{k}}^\dagger & a_{-\bm{k}}
\end{pmatrix}$ depending on the context. 
The magnon spectrum can be found by diagonalizing $H_{J}(\bm{k})$ by a matrix $T_{J}(\bm{k})$ that satisfies
\begin{equation}
T^\dagger_{J}(\bm{k}) H_{J}(\bm{k}) T_{J}(\bm{k})=\frac{1}{2}\tilde{\omega}^J_{\bm{k}},\quad T_{J}(\bm{k})^\dagger \tau_z T_{J}(\bm{k})=\tau_z, \label{eq.heisenberg_constraint}
\end{equation}
in which
\begin{equation}
\tilde{\omega}^J_{\bm{k}}=\begin{pmatrix}
\tilde{\omega}^J_{\bm{k},1} & \\
 & \tilde{\omega}^J_{\bm{k},-1}
\end{pmatrix}
\end{equation}
where $\tilde{\omega}^J_{\bm{k},\pm1}$ are non-negative.
Such a problem can be solved using the Colpa's method\cite{colpa1978diagonalization}, which is reviewed in Appendix~\ref{ap.BdG_review}.
We show the magnon spectrum of $H_J(\bm{k})$ (i.e. $\tilde{\omega}_{\bm{k},1}^J$)  in Fig.~\ref{fig.magnon_hsl} (a).

Let us note that although the magnetic order breaks the translation symmetry generated by $\bm{R}_1=(a,0)$ and $\bm{R}_2=(\frac{1}{2}a,\frac{\sqrt{3}}{2}a)$, where $a$ is the lattice constant, the Hamiltonian written in terms of HP operators respects the translation symmetry. 
This is because the magnetic ordering vector always rotates by $120^\circ$ counterclockwise (clockwise) about the global $z$-axis when translated by $\bm{R}_1$ ($\bm{R}_2$) while the terms quadratic in HP operators depend only on the cosine of the relative angle, as can be inferred from Eq.~\eqref{eq.s1dots2} [see also Appendix~\ref{ap.magnon_symmetry}]. Thus, we can still take the Bravais lattice generated by $\bm{R}_1$ and $\bm{R}_2$, and define $\bm{R}_3=\bm{R}_1-\bm{R}_2$. 
The reciprocal lattice vectors are then $\bm{G}_1=\frac{2\pi}{a}(1,-\frac{1}{\sqrt{3}})$, $\bm{G}_2=\frac{2\pi}{a}(0,\frac{2}{\sqrt{3}})$. 
The Hamiltonian also has a threefold rotation symmetry about the center of the yellow triangles in Fig.~\ref{fig.triangular_lattice} ($C_3$), a twofold rotation about the line through sites $1$ and $4$ ($C_{2y}'$), and a twofold rotation about the line through sites $1$ and $2$ ($C_{2x}$). These are the symmetries that are relevant for  gapless points between magnon and phonon bands, and their exact definitions are given in detail in Appendix~\ref{ap.magnon_symmetry}.

The magnon spectrum with just the Heisenberg interaction has three Goldstone modes\cite{chernyshev2009spin} at $\Gamma$, $K$, and $K'$. For the toy model, we will remove these Goldstone modes by adding easy-axis anisotropy along the direction of the magnetic ordering (local $z$-axis defined above),
\begin{equation}
{\cal H}_A=\sum_{i} A (S_{i}^{z})^2,
\end{equation}
where $A<0$. \footnote{This term is present in the $\textrm{RMnO}_3$, with R$=$Y, Lu \cite{oh2016spontaneous}. The magnon model in Ref. [\onlinecite{oh2016spontaneous}] also contains hard-axis anisotropy of the form $A'\sum_{i} (S_i^y)^2$ with $A'>0$. We ignore this term for simplicity. }
This removes all the Goldstone modes, but we will have to introduce an external magnetic field to remove the band degeneracies between magnon and phonon bands along $\Gamma K$ and $M\Gamma$, as explained in Sec.~\ref{ssec.mpctriangular}.

We can remove all the band degeneracies between magnons and phonons by applying external magnetic field along the global $z$-axis,
\begin{equation}
{\cal H}_{H}=\sum_i \vec{H}\cdot \vec{S}_i.
\end{equation}
This will tilt the magnetization direction towards the $z$-axis, which can be described by using mean field approximation\cite{elhajal2002symmetry}. 
Namely, let us assume that the spins will cant uniformly away from the plane \footnote{This may not be true for strong magnetic field which is evidenced by the $\frac{1}{3}$ magnetization plateau for Heisenberg antiferromagnet \cite{honecker2004magnetization}.}.
The energy per site is given by 
\begin{equation}
E=A S^2 \cos^2 \theta + H S \sin \theta+\frac{3}{2} J S^2  (2 \sin^2 \theta- \cos^2 \theta),
\end{equation}
where $\theta$ is the canting angle of the spin away from the 2D plane ($\theta>0$ corresponds to out-of-plane canting). By minimizing the energy, we obtain
\begin{equation}
\sin \theta = -\frac{H/S}{9J-2A}. \label{eq.canting_angle}
\end{equation}

If we perform the HP transformation for the full magnon Hamiltonian by taking into account the canting angle \footnote{The local axes in this case is shown in Fig.~\ref{fig.axes}},  we find 
\begin{align}
 \frac{A_{\bm{k}}}{S} =&-\frac{H}{S} \sin \theta+A (1-3 \cos^2\theta)- 6J(1-\cos^2 \theta+\tilde{\gamma_{\bm{k}}}), \nonumber \\
\frac{B_{\bm{k}}}{S}=&A\sin^2\theta+\frac{3J}{2}(1+2\cos^2\theta-2\sin^2\theta)\gamma_{\bm{k}}, \label{eq.magnon_components}
 \end{align}
 which are the coefficients of the magnon Hamiltonian defined in Eq.~\eqref{eq.heisenberg}. Here, we have defined $\tilde{\gamma}_{\bm{k}} =\frac{1}{12}\textrm{Re} [(1+\sin^2\theta-2\cos^2\theta+2i\sqrt{3}\sin \theta)(e^{i\bm{R}_{12}\cdot \bm{k}}+e^{i\bm{R}_{23}\cdot \bm{k}}+e^{i\bm{R}_{31}\cdot \bm{k}})]$. It can be checked that this formula reduces to the one defined previously if we turn off the anisotropy and magnetic field. 
 
The spectrum with the anisotropy and the magnetic field is shown in Fig.~\ref{fig.magnon_hsl} (b). If we assume that the land\'e g factor is about $1.6$, $H\sim 0.1$ meV corresponds to magnetic field of $1T$. We use the parameter $H=0.5$ meV, which would correspond to magnetic field of about $5$ T, and $A=-0.02$ meV.
Since there is only one magnon band, one cannot expect any topological band structure unless additional bosonic bands are taken into account. Moreover, the magnon Hamiltonian is real, so that the Berry curvature is zero.

\subsection{Phonon} \label{ssec.ptraingular}

Let us consider the phonon Hamiltonian for a triangular lattice. For simplicity, we will only consider the in-plane vibrations because there is no coupling between the out-of-plane vibration and the magnon in the approximation we use  [see Eq.~\eqref{eq.mp_coupling_model}]. The phonon Hamiltonian without magnetic field is given by
\begin{equation}
{\cal H}_p=\frac{1}{2}\sum_{\bm{R} \bm{R}'}\left[\bm{p}(\bm{R})^2\frac{1}{M}+\bm{u}(\bm{R}') K(\bm{R}'-\bm{R}) \bm{u}(\bm{R}) \right].
\end{equation}
Here, $\bm{R},\bm{R}'$ are the unit cell positions, $\bm{u}$ is the displacement, $\bm{p}$ is the momentum, and $K$ is the spring constant matrix. 
For simplicity, we only consider the longitudinal spring constant $\gamma$ for the nearest neighbors, which is typically several times larger than the transverse spring constant. For the spring constant matrix between sites $1$ and $2$, this can be done by taking 
\begin{equation}
K(\bm{R}_1)=\begin{pmatrix}
-\gamma & 0 \\
0 & 0
\end{pmatrix}.
\end{equation}
Due to the triangular lattice symmetry of phonon, which we review in Appendix~\ref{ap.phonon_symmetry}, we have $K(\bm{R}_2)=C_6 K(\bm{R}_1) C_6^{-1}$ and $K(\bm{R}_3)=C_3 K(\bm{R}_1) C_3^{-1}$. Finally, $K(\bm{R}=0)=\textrm{diag}(3\gamma, 3\gamma$) follows from the constraint that $\sum_{\bm{R}} K(\bm{R})=0$. The phonon Hamiltonian thus constructed is naturally symmetric with respect to $C_{6z}$, $C_{2x}$, and $C_{2y}'$ symmetries.

\begin{figure}[t]
\centering
\includegraphics[width=8.5cm]{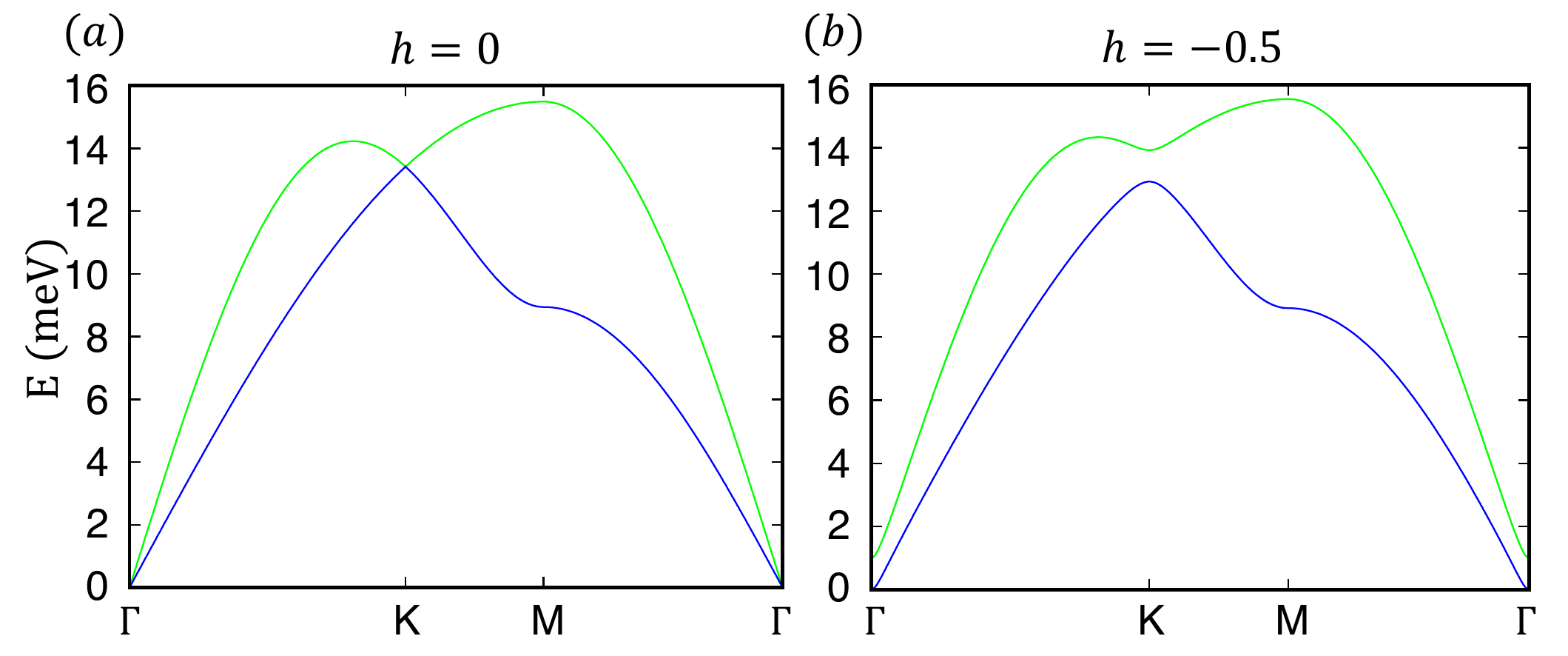}
\caption{Phonon spectrum along the high symmetry line in the unit of meV. (a) The phonon spectrum with $\hbar^2 \gamma/M=40$ $\textrm{(meV)}^2$ and no effective magnetic field. (b) The phonon spectrum with effective magnetic field $h \hbar=-0.5$ meV.
}
\label{fig.phonon_hsl}
\end{figure}

The dynamical matrix is defined to be
\begin{equation}
D(\bm{k})=\sum_{\bm{R} }{\frac{1}{M}K(\bm{R})e^{i\bm{k}\cdot\bm{R}}} ,\label{eq.dynamical_matrix}
\end{equation}
where $M$ is the mass of the ion, $D_{xx} (\bm{k})= \frac{\gamma}{M} (3-2\cos k_x - \cos \frac{k_x}{2} \cos \frac{\sqrt{3}k_y}{2} )$, $D_{xy}(\bm{k}) = D_{yx}(\bm{k}) =\frac{\gamma}{M}\sqrt{3} \sin \frac{k_x}{2} \sin \frac{\sqrt{3}k_y}{2}$, $D_{yy}(\bm{k})=3 \frac{\gamma}{M} (1-\cos \frac{k_x}{2} \cos \frac{\sqrt{3}k_y}{2})$.
The resulting phonon band structure is shown in Fig.~\ref{fig.phonon_hsl} (a).
We see that there are two acoustic phonon bands which are degenerate at $\Gamma$ and $K$.

To lift the degeneracy between the phonon bands, we can introduce external magnetic field. 
We review the details of how this can be done in Sec.~\ref{ssec.phonon_review}.
For our purposes, it suffices to note that the phonon Hamiltonian with magnetic field can be written as
\begin{equation}
{\cal H}_p=\sum_{\bm{k}}\bm{x}^\dagger(\bm{k}) H_p(\bm{k}) \bm{x}(\bm{k}). 
\end{equation}
Here, we have redefined $\bm{p}(\bm{k})\rightarrow \sqrt{M}\bm{p}(\bm{k})$ and $\bm{u}(\bm{k})\rightarrow \bm{u}(\bm{k})/\sqrt{M}$, and defined the operator
\begin{equation}
\bm{x}(\bm{k})=\begin{pmatrix}
\bm{p} (\bm{k}) \\
\bm{u}(\bm{k})
\end{pmatrix}, 
\end{equation}
and the matrices
\begin{equation}
 H_{p}(\bm{k})=\frac{1}{2}\begin{pmatrix}
I_{sd} & -A \\
A & D(\bm{k})
\end{pmatrix},
\quad
A=\begin{pmatrix}
0 & h \\
-h & 0
\end{pmatrix}\label{eq.phonon_hamiltonian_t}.
\end{equation}
The parameter $h$ contains the coupling between phonon and magnetization.

The energy spectrum can be found by solving the Hamiltonian equation of motion [see Eq.~\eqref{eq.phonon_eigenvector}]. 
The phonon spectrum with magnetic field is shown in Fig.~\ref{fig.phonon_hsl}  (b). We note that the energy scale associated with the magnetic field is $h \hbar \sim 0.002$ meV for magnetic field about $1T$ at 5.45K\cite{sheng2006theory,zhang2011phonon} for a paramagnet,  which is quite small. We will put $h \hbar=-0.5$ meV to clarify the role of magnon-phonon interaction.

\subsection{Magnon-phonon coupling} \label{ssec.mpctriangular}

Let us use the exchange magnetostriction model for the magnon-phonon coupling \cite{kittel1960model},
\begin{equation}
{\cal H}_{c}=\sum_{\langle ij\rangle} K_{mp} \bm{R}_{ij}\cdot \Delta \bm{u}_{ij} \bm{S}_i\cdot \bm{S}_j , \label{eq.mp_coupling_model}
\end{equation}
where $\bm{R}_{ij}=\frac{1}{a}(\bm{R}_i-\bm{R}_j)$ is a unit vector and $\Delta\bm{u}=\bm{u}_i-\bm{u}_j$. This form of the Hamiltonian can be obtained from the Heisenberg model by assuming that the exchange integral $J$ depends on the distance between the atoms, $J(|\bm{r}_i-\bm{r}_j|)\approx J+K_{mp} \bm{R}_{ij} \cdot \Delta \bm{u}_{ij}$ where $\bm{r}_i=\bm{R}_i+\bm{u}_i$. Note that out-of-plane vibration will not couple to magnons in this model. For a non-collinear antiferromagnet, magnon-phonon coupling can arise naturally in quadratic order because $\bm{S}_i\cdot \bm{S}_j$ contains terms linear in the HP operators. 

In Sec.~\ref{ssec.BdG}, we will discuss  two methods to solve the magnon-phonon coupling problem: we can work either with 
\begin{equation}
\Phi_{\bm{k}}=(a_{\bm{k}},a^{\dagger}_{-\bm{k}}, \bm{p}^{T}_{\bm{k}},\bm{u}^{T}_{\bm{k}} ),
\end{equation}
where $\bm{u}(\bm{k})$ is the displacement and $\bm{p} (\bm{k})$ is the conjugate momentum in the Fourier space, or with 
\begin{equation}
\Psi_{\bm{k}}=(a_{\bm{k}},b_{1,\bm{k}},b_{2,\bm{k}},a^{\dagger}_{-\bm{k}}, b^{\dagger}_{1,-\bm{k}},b^{\dagger}_{2,-\bm{k}} ),
\end{equation}
where $b_{1,\bm{k}}$ and $b_{2,\bm{k}}$ are the phonon operators in the Fourier space. Although the energy spectrum of these two methods are the same, their Berry curvature will be different. In order to calculate the thermal Hall conductivity, the correct Berry curvature is computed by working in $\Phi_{\bm{k}}$ basis. In order to compare these two methods, we will present the results using both methods.

Let us first work in the $\Psi_{\bm{k}}$ basis. 
It can be shown that up to terms linear in the HP operators, 
\begin{equation}
\bm{S}_1\cdot \bm{S}_j=c a_1+c^* a_1^\dagger-c^* a_2 - c a_2^\dagger, ~j=2,10,12,
\label{eq.s1s2_to_magnon}\end{equation}
where $c=\frac{S^{3/2}}{2} \sqrt{\frac{3}{2}} \cos \theta - \frac{3 S^{3/2}\cos \theta \sin \theta}{2\sqrt{2}}i$.
Similarly, we have
\begin{equation}
\bm{S}_1\cdot \bm{S}_j=-c^* a_1-c a_1^\dagger+c a_3+c^* c_3^\dagger,~ j=3,4,11.
\label{eq.s1s3_to_magnon}
\end{equation}
This pattern arises from the difference in the ordering direction of $j=2,10,12$ and $j=3,4,11$ with respect to the spin at site $1$.

Let us now note that the Hamiltonian for the magnon-phonon coupled system also has the translation symmetry of the underlying triangular lattice. This is because $\bm{S}_i\cdot \bm{S}_j$ only depends on whether the direction of $\langle \bm{S}_{j}\rangle$ (classical spin direction) is rotated clockwise or counterclockwise by $120^\circ$ about the global $z$-axis compared to $\langle \bm{S}_{i}\rangle$, as can be seen from Eqs.~\eqref{eq.s1s2_to_magnon},~\eqref{eq.s1s3_to_magnon}. 
Taking the Fourier transform by taking into account the translation symmetry, we obtain the following contribution to the magnetoelastic Hamiltonian from translations of the magnon-phonon coupling between the pair $(1,2)$ \footnote{Here, we have not written terms that cancel after adding the contribution from other pairs}:
\begin{align}
{\cal H}_{c}^{(1,2)}=&\sum_{\sigma=1,2,\bm{k}} 2\sqrt{\frac{\hbar}{M}}K_{mp} \bm{R}_{12} \cdot [\bm{\epsilon}_{\sigma}(\bm{k}) b_{\bm{k},\sigma} +\bm{\epsilon}_{\sigma}^*(-\bm{k}) b_{\bm{k},\sigma}^\dagger] \nonumber \\
&\times[\textrm{Re}(-ce^{-i\bm{k}\cdot \bm{R}_{12}}) a_{-\bm{k}}+\textrm{Re}(-c^*e^{-i\bm{k}\cdot \bm{R}_{12}}) a^\dagger_{\bm{k}}], \label{eq.triangular_mpc1}
\end{align}
where $\bm{\epsilon}_{\sigma}(\bm{k})$ and $b_{\bm{k},\sigma}$ are the polarization vector and the phonon operator defined in Sec.~\ref{ssec.phonon_review}. 
The contribution from the other bonds can be found by permuting the indices of $\bm{R}_{ij}$. For the pair $(1,10)$, ${\cal H}_{c}^{(1,10)}$  is obtained by permuting $(1,2)\rightarrow (2,3)$, and ${\cal H}_{c}^{(1,12)}$ is obtained by permuting  $(2,3)\rightarrow (3,1)$.  Therefore, the total magnon-phonon coupling Hamiltonian is  ${\cal H}_{c}={\cal H}_{c}^{(1,2)}+{\cal H}_{c}^{(1,10)}+{\cal H}_{c}^{(1,12)}$

Collecting the magnon, phonon, and the magnon-phonon coupling Hamiltonian, we can write
\begin{equation}
H=\sum_{\bm{k}} \Psi_{\bm{k}}^\dagger \tilde{H}_{me}(\bm{k}) \Psi_{\bm{k}}. \label{eq.triangular_hme_tilde}
\end{equation}
 The matrix $\tilde{H}_{me}(\bm{k})$ has the bosonic BdG form because the magnon and phonon operators satisfy the bosonic canonical commutation relation. Thus, the Hamiltonian can be diagonalized by using the Colpa's method \cite{colpa1978diagonalization}, which is reviewed in Appendix \ref{ap.BdG_review}.

\begin{figure}[t]
\centering
\includegraphics[width=8cm]{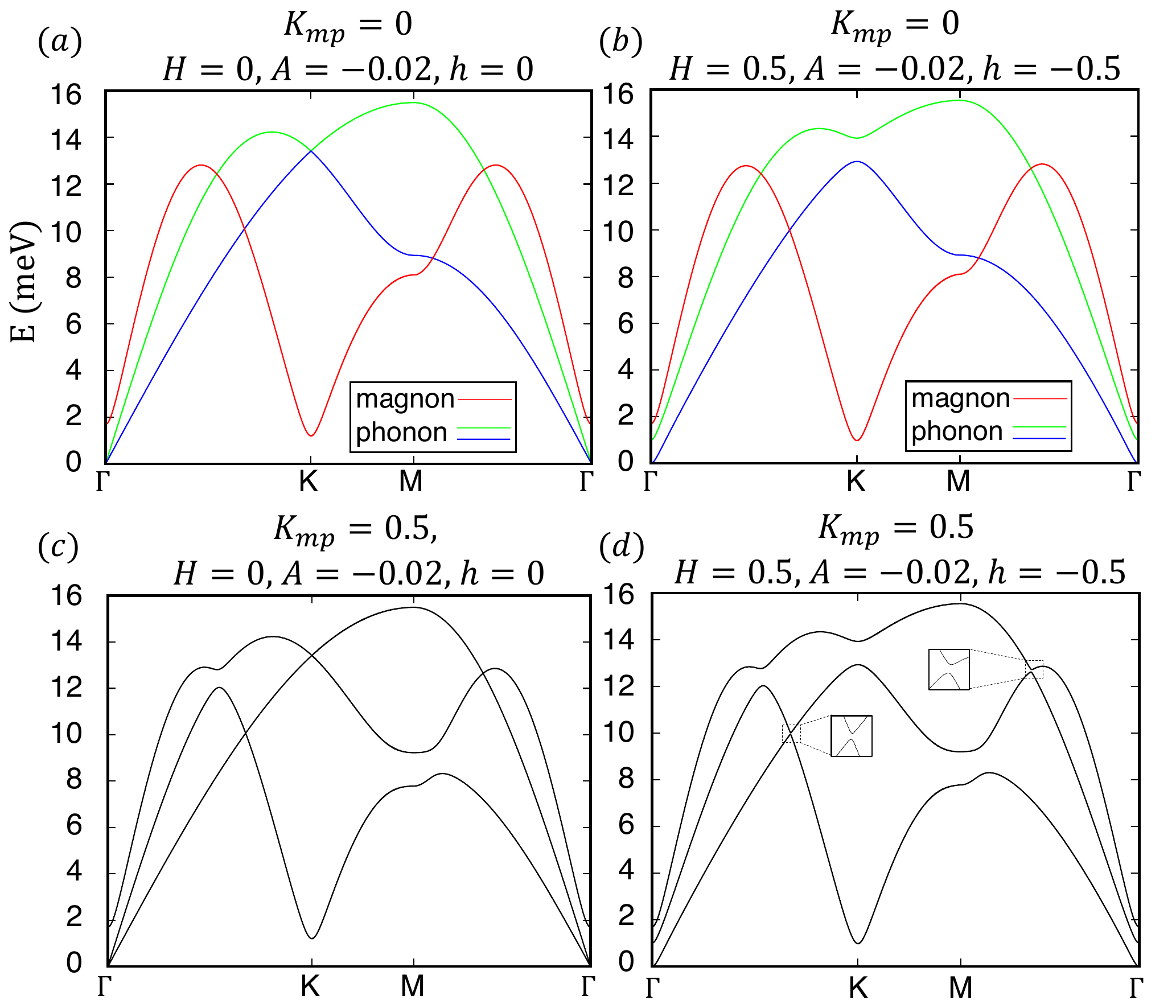}
\caption{Influence of magnon-phonon coupling ($K_{mp}$) and magnetic field ($H$ and $h$) on the magnon phonon band structure. (a) Magnon and phonon bands without magnon-phonon coupling ($K_{mp}=0$) and without magnetic field ($H=h=0$). (b) Magnon and phonon bands without magnon-phonon coupling ($K_{mp}=0$) and with magnetic field ($H\neq 0$, $h\neq 0$). (c) Magnon and phonon bands with magnon-phonon coupling ($K_{mp}\neq 0$) and without magnetic field ($H=h=0$). Note that the gap along $\Gamma K$ and $M\Gamma$ does not open. (d) Magnon and phonon bands with magnon-phonon coupling ($K_{mp}\neq 0$) and with magnetic field ($H\neq0$, $h\neq 0$). Note that the gap along $\Gamma K$ and $M\Gamma $ opens. If we put $\hbar=1$ and measure energy in units of meV, (d) can be reproduced by putting $J=2$, $S=2$, $H=0.5$, $A=-0.02$, $K_{mp}\sqrt{\frac{\hbar}{M}}=0.5$, $\gamma/M=40$, and $h=-0.5$.} 
\label{fig.mp_hsl}
\end{figure}

The spectrum of $\tilde{H}_{me}(\bm{k})$ without magnetic field is shown in Fig.~\ref{fig.mp_hsl} (a) and (c). We have plotted the magnon and phonon spectrum without magnon-phonon coupling ($K_{mp}=0$) in Fig.~\ref{fig.mp_hsl} (a) to compare with the case with magnon-phonon coupling ($K_{mp}\neq0$) in Fig.~\ref{fig.mp_hsl} (c).
The strength of magnon-phonon coupling, $K_{mp}\sqrt{\frac{ f \hbar}{M}}$ where $f=10^3\times \hbar/e \approx 0.658 \times 10^{-12}$ , can be expected to be about $0.3$ $[\textrm{meV}/\textrm{s}^{1/2}]$ \cite{oh2016spontaneous}. The numerical factor $f$ arises naturally if we take $\hbar=1$ and use 1 meV as the unit of energy for magnon-phonon coupling problem. We use a reasonable value of $0.5$ $[\textrm{meV}/\textrm{s}^{1/2}]$ for our model.

Let us notice that the gap does not open up along the high symmetry lines $\Gamma K$ and $M\Gamma$ even in the presence of the magnon-phonon coupling. 
This is because of the $C_{2x}$ and $C_{2y}'$  symmetries mentioned previously. These two symmetries are present in the magnon-phonon coupling Hamiltonian as well, as explained in more detail in Appendix~\ref{ap.mp_symmetry}, and are therefore relevant for determining whether magnon bands and phonon bands can hybridize along the high symmetry lines. 
For the magnon band,  the $C_{2x}$ eigenvalue is $1$ along the $\Gamma K$ line,  and the  $C_{2y}'$ eigenvalue is $-1$ along the $M\Gamma$ line. 
For the phonon bands, the $C_{2x}$ eigenvalue along the $\Gamma K$ line for the band with higher (lower) energy is $+1$ ($-1$), and the  $C_{2y}'$ eigenvalue along the $M \Gamma$ line for the band with higher (lower) energy is $+1$ ($-1$). 
Because energy bands with the same (different) eigenvalues can (cannot) hybridize, the gap closing points between magnon and phonon bands that remain along the high symmetry lines in Fig.~\ref{fig.mp_hsl} (c) can be explained.

When the external magnetic field is turned on, the phonon Hamiltonian does not have the $C_{2x}$ and $C_{2y}'$ symmetry because the effective Lorentz force an ion will feel when moving in the positive $y$ direction is not the same as when it is moving in the negative $y$ direction. 
Thus, we should expect that the gap will open. 
This is shown in Fig.~\ref{fig.mp_hsl} (b) and (d), where we have drawn the magnon and phonon spectrum with magnetic field ($h\neq 0,H\neq 0$) and without magnon-phonon coupling ($K_{mp}=0$) in (b) for comparison with the case when there is magnon-phonon coupling ($K_{mp}\neq 0$) in (d).  
We see that both the magnetoelastic coupling and the external magnetic field are necessary to fully open the gap between the magnon-polaron bands.

We have mentioned that the magnetoelastic Hamiltonian can also be written in basis $\Phi_{\bm{k}}$, 
\begin{equation}
H=\sum_{\bm{k}} \Phi_{\bm{k}}^\dagger H_{me}(\bm{k}) \Phi_{\bm{k}}, \label{eq.triangular_hme}
\end{equation}
The details of this method can be found in Sec.~\ref{ssec.diag_me}. Here, we only mention that $H_{me}(\bm{k})$ takes the form in Eq.~\eqref{eq.magnetoelastic_hamiltonian}, and that the $2\times 4$ magnon-phonon coupling Hamiltonian $H_{c}(\bm{k})$ is given by
\begin{equation}
H_c(\bm{k})=\begin{pmatrix}
0 & 0 &  \bm{v}(\bm{k}) \\
0 & 0 & \bm{v}(-\bm{k})
\end{pmatrix}, \label{eq.triangular_mpc2}
\end{equation}
where $\bm{v}(\bm{k})$ is the $1\times 2$ column vector given by $\bm{v}(\bm{k})=K_{mp} \sqrt{\frac{\hbar}{M}} \textrm{Re} [-c e^{i \bm{k}\cdot \bm{R}_{12}} \bm{R}_{12} +(12\leftrightarrow 23)+(12\leftrightarrow 31)]$. After performing the transformation to bosonic BdG Hamiltonian $H_{s}$ as in Eq.~\eqref{eq.simplified_hamiltonian}, we can use Colpa's method to diagonalize the Hamiltonian.

\section{Berry Curvature and Thermal Hall Conductivity in Triangular Antiferromagnet} \label{sec.bc_tc}

In this section, we compare the Berry curvature computed from $\tilde{H}_{me}(\bm{k})$ and $H_{me}(\bm{k})$, and compute the thermal Hall conductivity of the model presented in Sec.~\ref{sec.mptriangular}.
Large Berry curvature is induced in the anticrossing regions because of the magnon-phonon coupling and the effective magnetic field in phonon.
This renormalize the thermal Hall conductivity arising from phonons. 
We also show that the decoupled magnon-polaron bands are topological.

\subsection{Berry Curvature} \label{ssec.triangularBerry}

\begin{figure}[ht]
\centering
\includegraphics[width=8cm]{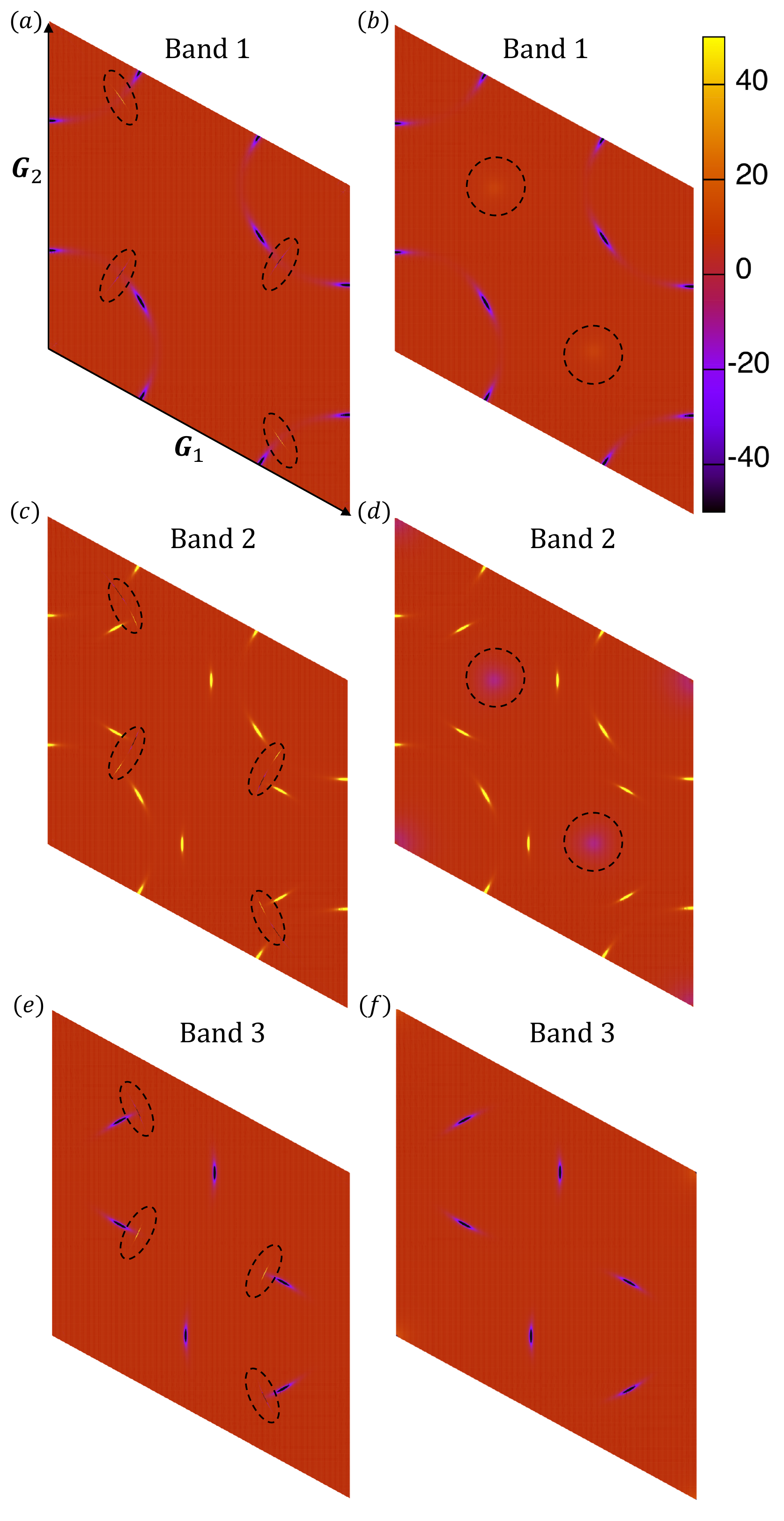}
\caption{Berry curvature density of magnon-phonon hybridized bands in the first Brillouin zone with the lattice constant $a=1$. The energy bands are labelled $1,2,3$ from highest to lowest energy and carry Chern numbers $-2$, $4$, $-2$ respectively. (a) and (b) shows the Berry curvature density for energy band $1$ using the $\tilde{H}_{me}(\bm{k})$  and $H_{me}(\bm{k})$, respectively. (c) and (d) are similar plots for energy band $2$, and (e) and (f) are similar plots for energy band $3$.}
\label{fig.mp_berry}
\end{figure}

\begin{figure}[t]
\centering
\includegraphics[width=8.5cm]{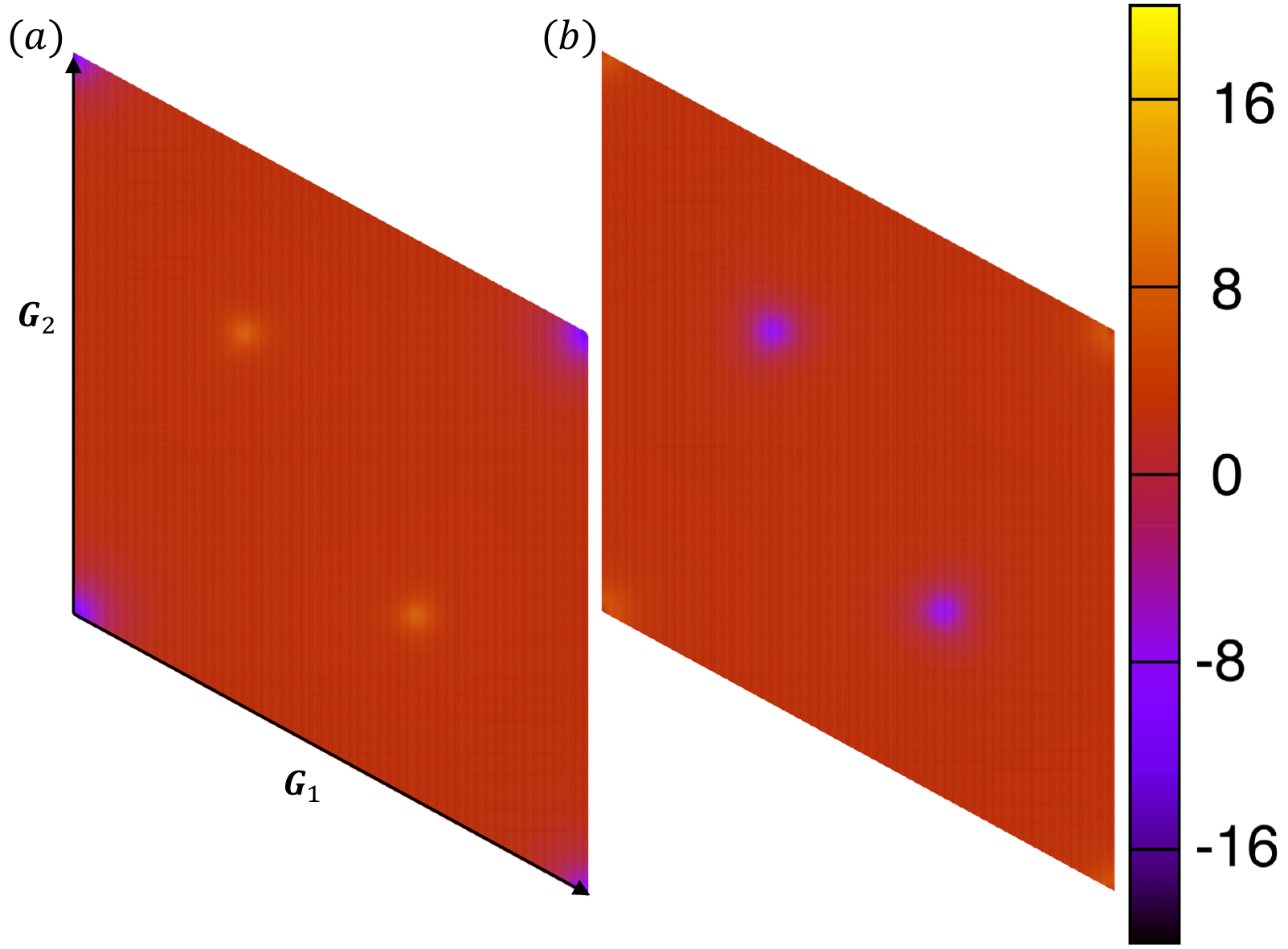}
\caption{Berry curvature density of phonon in the first Brillouin zone with the lattice constant $a=1$ and with the same parameters used in Fig.~\ref{fig.phonon_hsl} (b). (a) and (b) are the plots for the higher and lower energy band, respectively. }
\label{fig.p_berry}
\end{figure}

Since the magnetoelastic bands are fully gapped in the presence of magnetic field and magnetoelastic coupling, each hybridized band can carry quantized Chern number.
In Fig.~\ref{fig.mp_berry}, we compare the Berry curvature calculated for the magnon-phonon coupled band with $\tilde{H}_{me}(\bm{k})$ (defined in Eqs.~\eqref{eq.triangular_hme_tilde} and \eqref{eq.magnon_phonon_hamiltonian}) and $H_{me}(\bm{k})$ (defined in Eqs.~\eqref{eq.triangular_hme} and \eqref{eq.magnetoelastic_hamiltonian}). The Berry curvature density computed using $\tilde{H}_{me}(\bm{k})$ is shown in Figs.~\ref{fig.mp_berry} (a), (c), (e), and it should be compared with that computed using $H_{me}(\bm{k})$ which is shown in Figs.~\ref{fig.mp_berry} (b),(d),(f). 
The most noticeable difference is that the contribution from phonon Berry curvature, which is shown in Fig.~\ref{fig.p_berry}, can be seen in Fig.~\ref{fig.mp_berry} (b) and (d) [indicated with dotted circle], but not in (a) and (c). Another difference is that the Berry curvature computed from $\tilde{H}_{me}(\bm{k})$ shows spots of large Berry curvature, indicated by dotted ellipse in Fig.~\ref{fig.mp_berry} (a),(c),(e), which are not present in Figs.~\ref{fig.mp_berry}(b),(d),(f) computed from $H_{me}(\bm{k})$. Their origin can be traced back to the fact that $\tilde{H}_{me}(\bm{k})$ is not smooth when computed numerically. This is because $\tilde{H}_{me}(\bm{k})$ depends on $\bm{\epsilon}_{\sigma}(\bm{k})$ whose phase factor is indeterminate. This implies that the Berry phase computed from $\tilde{H}_{me}(\bm{k})$ does not behave well for numerical computation. We explain this in detail in Appendix~\ref{ap.berry}

After integration of the Berry curvature, we find that the Chern numbers of the bands, from highest to lowest energy, are $-2$, $4$, $-2$ respectively.
Surprisingly, the Chern numbers computed from $H_{me}(\bm{k})$ and $\tilde{H}_{me}(\bm{k})$ are equivalent. 
In general, however, we should not expect the Chern numbers computed with these two methods to be equivalent.

Finally, let us note that the effective magnetic field $h$ in phonon is essential for the presence of Berry curvature. 
To see this, let us first notice that $H_{me}(\bm{k})$ is real because $H_m(\bm{k})$, $H_p(\bm{k})$, and $H_c(\bm{k})$ are real [see Eqs.~\eqref{eq.magnon_components},\eqref{eq.phonon_hamiltonian_t}, and \eqref{eq.triangular_mpc2}].
When $h=0$, this guarantees that the Berry curvature vanishes, which we show in Appendix~\ref{ap.berry_reality_condition}. 
Therefore, although the magnon-phonon coupling does not by itself induce Berry curvature, it can induce large Berry curvature in the presence magnetic field in phonon Hamiltonian.
The same conclusion holds for the Berry curvature computed from $\tilde{H}_{me}(\bm{k})$. This is because if $h=0$, we can choose $\bm{\epsilon}_{\sigma}(\bm{k})$ to be real. 
Then, it is immediate from Eq.~\eqref{eq.triangular_mpc1} that the magnon-phonon coupling terms are real, so that $\tilde{H}_{me}(\bm{k})$ is real. It follows from this that there is no Berry curvature.

\subsection{Thermal Hall Conductivity} \label{ssec.thermal}

\begin{figure}[ht]
\centering
\includegraphics[width=8cm]{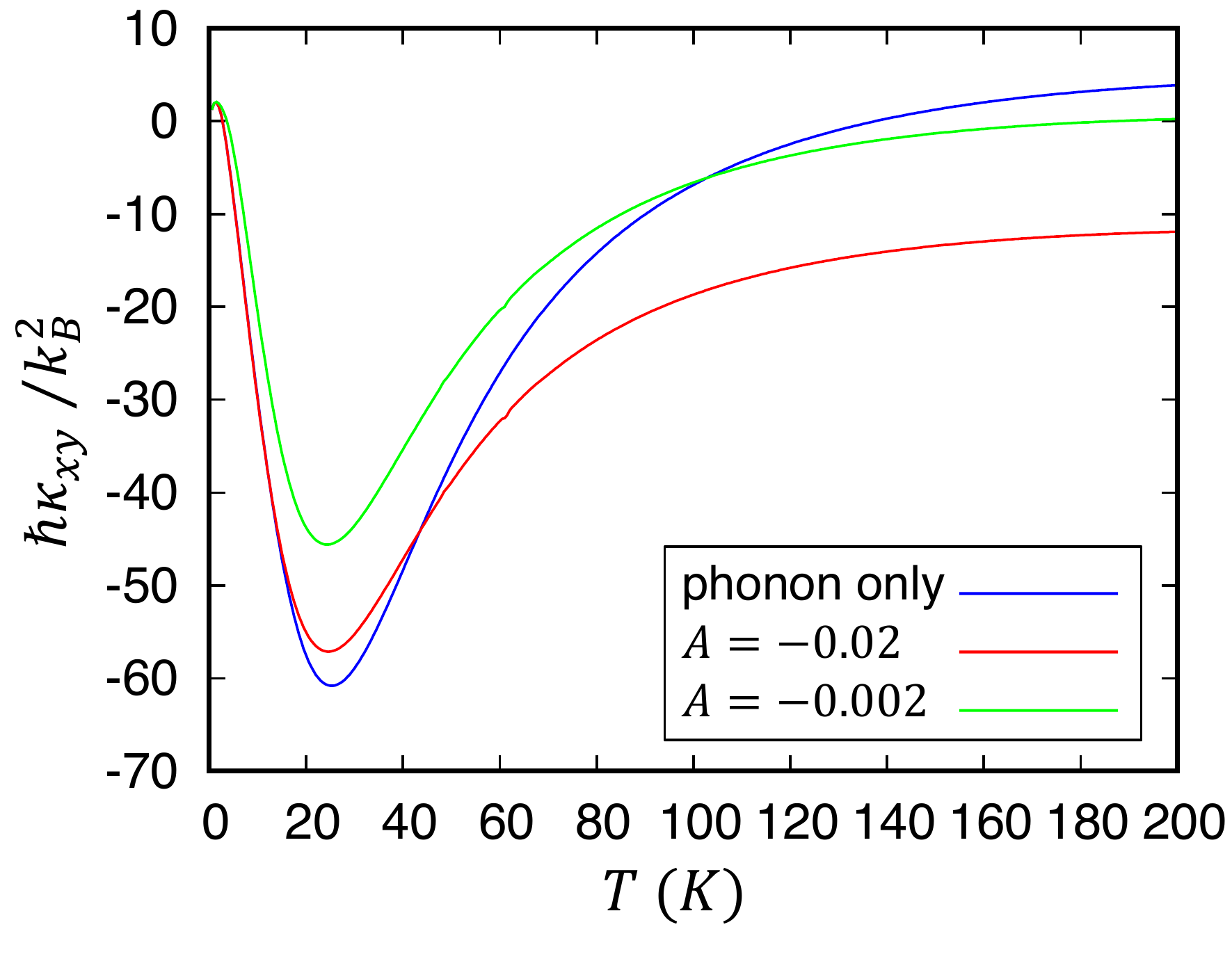}
\caption{Thermal Hall conductivity $\kappa_{xy}$. The blue line is $\kappa_{xy}$ for phonon with parameters given in Fig.~\ref{fig.phonon_hsl}.  The red line is $\kappa_{xy}$ for magnon-polaron with parameters given in Fig.~\ref{fig.mp_hsl}. The green line is computed with the same parameters except the easy-axis anisotropy, which is reduced to $A=-0.002$. The $x$ axis is temperature in Kelvins and the $y$-axis is the dimensionless thermal Hall conductivity, $\hbar \kappa_{xy}/k_B^2$.}
\label{fig.thermal_Hall}
\end{figure}

The formula for thermal Hall conductivity can be derived by either semi-classical theory or linear response theory. 
For non-BdG bosonic Hamiltonian, it was shown that the two approaches are equivalent \cite{matsumoto2011theoretical,matsumoto2011rotational}.
This equivalence holds even for BdG Hamiltonian, as we now show.
The formula for the thermal Hall conductivity derived by using semi-classical wave-packet approach is given by \cite{zhang2016berry}
\begin{equation}
\kappa_{xy}=\frac{1}{2\hbar TV}\sideset{}{'}\sum_{n=-N}^{N}\sum_{\bm{k}} \Omega_{n,\bm{k}}\int_{\hbar \omega_{n,\bm{k}}}^{\infty} E^2 \frac{\partial g}{\partial E}dE ,\label{eq.thermal_semiclassical}
\end{equation}
where $\Omega_{n,\bm{k}}$ is the $z$-component of the Berry curvature computed as in Eq.~\eqref{eq.berry_curvature}, $\hbar \omega_{\bm{k},n}$ is the energy with $\hbar \omega_{\bm{k},n}>0$ ($<0$) for $n>0$ ($<0$),  $g(E)=\frac{1}{e^{E/k_BT}-1}$ is the Bose-Einstein distribution, and the $ '$ indicates that there is no summation over $n=0$. Let us note the formula
\begin{equation}
\int_{\hbar \omega_{n,\bm{k}}}^{\infty}E^2 \frac{\partial g}{\partial E} dE=-k_B^2 T^2 c_{2}\big((g(\hbar \omega_{n,\bm{k}})\big),
\end{equation}
where $c_{2}(x)\equiv \int_{0}^{x}[\log(1+\rho^{-1})]^2$. This can be derived by making the substitution $\rho=g$, so that $E=k_{B}T \log(1+\rho^{-1})$. Combining this with the properties $\Omega_{-n,\bm{k}}=-\Omega_{-n,\bm{k}}$, $\omega_{-n,\bm{k}}=-\omega_{n,\bm{k}}$, $\frac{\partial g(-E)}{dE}=-\frac{\partial g(E)}{dE}$ and $c_{2}(\infty)=\frac{\pi^2}{3}$, we can convert the summation for $n<0$ to summation over $n>0$. After a short calculation, we arrive at the following expression for the thermal Hall conductivity, which can also be derived from the linear response theory \cite{qin2012berry,matsumoto2014thermal}
\begin{equation}
\kappa_{xy}=-\frac{k_B^2 T}{V\hbar }\sum_{\bm{k}}\sum_{n=1}^{N} \left[c_2(g(\hbar \omega_{n,\bm{k}}))-\frac{\pi^2}{3}\right] \Omega_{n,\bm{k}}.
\end{equation}
We thus see that the thermal Hall conductivity for BdG Hamiltonian derived from semi-classical theory agrees with that derived from linear response theory.


We show the thermal Hall conductivity of magnon-polaron as a function of temperature with red line in Fig.~\ref{fig.thermal_Hall}, calculated using the  parameters used in Fig.~\ref{fig.mp_hsl} (d). As a comparison, we plot the case without magnon-phonon coupling with blue line, which is equal to the phonon Hall conductivity because the magnon Hall conductivity vanish.  The Berry curvature arising from magnon-phonon interaction contributes noticeably to the thermal Hall conductivity at high temperature. This is because the hybridization between magnon and phonon occurs significantly only at high energies. On the other hand, if we reduce the anisotropy from $a=-0.02$ to $-0.002$ meV, magnon and phonon can hybridize significantly also at lower energies, and this is accompanied by a topological phase transition with the Chern numbers given by $0$, $2$, $-2$, from bands with highest energy to lowest energy. The thermal Hall conductivity for this case is shown in green line, and we see that the thermal Hall conductivity is now significantly renormalized at lower temperatures.

\section{Diagonalization of Magnetoelastic Hamiltonian} \label{sec.diagonalization}

In this section, we clarify the relation between the magnetoelastic Hamiltonian and the BdG Hamiltonian. To introduce the notations used for phonons, we begin with a brief review of the theory of phonon in a two dimensional lattice with net out-of-plane magnetization, which couples to phonons through the Raman interaction \cite{zhang2011phonon,zhang2010topological,qin2012berry}. Then, we clarify the relation between the phonon Hamiltonian to the BdG Hamiltonian. Using this, we then present a method to diagonalize the  magnetoelastic Hamiltonian without introducing the phonon operators, based on Colpa's method of diagonalizing bosonic BdG Hamiltonian, which is reviewed in Appendix~\ref{ap.BdG_review}. For complicated systems, this can simplify the work involved in solving the hybridization problem. We will always put $\hbar=1$.

\subsection{Review of phonon Hamiltonian in effective magnetic field} \label{ssec.phonon_review}

The Hamiltonian of an ion moving in a static out-of-plane magnetic field $\bm{B}$ can be written by making the substitution $\bm{p} \rightarrow \bm{p}-q\bm{A}$ where $\bm{p}$ is the momentum conjugate to the displacement $\bm{u}$, $q$ is the charge of the ion, and $\bm{A}=\frac{1}{2}\bm{B} \times \bm{u}$ is the vector potential. Then, the kinetic part of the Hamiltonian is 
\begin{equation}
\frac{1}{2m}\left|\bm{p}-\frac{q}{2}\bm{B}\times \bm{u}\right|^2=\frac{1}{2m}\left|\bm{p}-\begin{pmatrix}
0 & -\frac{qB}{2} \\
\frac{qB}{2} & 0
\end{pmatrix}\bm{u}\right|^2.
\end{equation}

The effective Hamiltonian of lattice vibration in the presence of magnetization can be written in a similar way.
Let $\bm{u}_\alpha (\bm{R})$ denotes the two-dimensional displacement vector of an ion multiplied by the square root its mass, $m_\alpha$. Here, $\bm{R}$ is the unit cell position and  $\alpha$ is the sublattice index. Similarly, let $\bm{p}_\alpha (\bm{R})$ be the conjugate momentum divided by  square root of the mass. We will denote the charge of ion $\alpha$ by $q_\alpha$, $\alpha=1,...,s$, where $s$ is the number of sublattice. We will often omit the sublattice index and write, for example, $\bm{u}$ to mean $(\bm{u}_1,...,\bm{u}_s )$. 
The phonon Hamiltonian is given by\cite{zhang2010topological,holz1972phonons} 
\begin{widetext}
\begin{equation}
{\cal H}_{p}=\frac{1}{2}\sum_{\alpha \beta \bm{R} \bm{R}'} \left[\left\{ \bm{p}_{\alpha}(\bm{R})^2+2 \bm{u}_{\alpha}(\bm{R})A_{\alpha \alpha} \bm{p}_{\alpha}(\bm{R})\right\} \delta_{\alpha \beta}
\delta_{\bm{R},\bm{R}'}+\bm{u}_\alpha(\bm{R})\left\{K_{\alpha \beta}(\bm{R}-\bm{R}')-(A^2)_{\alpha \beta}\right\}\bm{u}_{\beta}(\bm{R}')\right]. \label{eq.phononHR}
\end{equation}
\end{widetext}
Here, $A$ is a block diagonal matrix with blocks $A_{\alpha \beta}=\delta_{\alpha \beta}\Lambda_{\alpha}$, where $\Lambda_{\alpha}$ is a $d\times d$ matrix and $d$ is the spatial dimension, which is $2$ for the present case. This matrix contains the coupling between the ions and the effective magnetic field
\begin{align}
\Lambda_\alpha=\begin{pmatrix}
0 & h \\
-h & 0
\end{pmatrix},
\label{eq.Lambda}\end{align}
where we have defined $h_\alpha=-q_\alpha B/2m_\alpha$. 
As we have mentioned before, $B$ is the effective magnetic field, which is proportional to the local magnetization in the $z$ direction. This coupling between $\bm{u}$ and $\bm{p}$ can occur by the Raman-type interaction of the form $g \bm{M}\cdot(\bm{u}\times\bm{p})$, where $\bm{M}$ is the average magnetization \cite{ioselevich1995strongly}.

We use the following convention for the Fourier transformation of phonons:
\begin{equation}
\bm{u}_{\alpha}(\bm{R})=\frac{1}{\sqrt{{\cal N}}}\sum_{\bm{R}} \bm{u}_\alpha(\bm{k}) e^{i(\bm{R}+\bm{\delta}_\alpha)\cdot \bm{k}}.
\end{equation}
Here, $\bm{R}$ is the position of the cell, $\delta_\alpha$ is the displacement from $\bm{R}$ to the equilibrium position of the atom in that cell, and ${\cal N}$ is the total number of unit cells. 
The Hamiltonian after the Fourier transformation is given by \cite{zhang2010topological}
\begin{widetext}
\begin{align}
{\cal H}_p&=\frac{1}{2}\sum_{\alpha \beta k} \left[\left\{\bm{p}_\alpha(-\bm{k}) \cdot \bm{p}_\alpha(\bm{k}) +2 \bm{u}_{\alpha}(-\bm{k})A_{\alpha \alpha} \bm{p}_\alpha(\bm{k})\right\}\delta_{\alpha \beta}+\bm{u}_{\alpha}(-\bm{k})D_{\alpha \beta}(\bm{k})\bm{u}_\beta(\bm{k})\right] \nonumber \\
&=\sum_{\bm{k}}
\bm{x}(-\bm{k})^{T} H_{\textrm{p}}(\bm{k}) \bm{x}(\bm{k})
\end{align}
\end{widetext}
where
\begin{equation}
D_{\alpha \beta}(\bm{k})=-(A^2)_{\alpha \beta}+\sum_{\Delta \bm{R}, \alpha \beta} K_{\alpha \beta}(\Delta \bm{R})  e^{i \bm{k}(\Delta \bm{R}+\bm{\delta}_\alpha-\bm{\delta}_\beta)},  \label{eq.dynamical matrix}
\end{equation} $\bm{x}(\bm{k})=(\bm{p}(\bm{k}),\bm{u}(\bm{k}))$, and 
\begin{equation}
H_{p}(\bm{k})=\frac{1}{2}\begin{pmatrix}
I_{sd} & -A \\
A & D(\bm{k})
\end{pmatrix}, \label{eq.phonon_hamiltonian}
\end{equation}

By writing the Hamilton's equations of motion for $\bm{u}(\bm{k})$ and $\bm{p}(\bm{k})$, we see that the eigenvalue problem that must be solved is
\begin{equation}
H_{\textrm{eff}} (\bm{k})\bm{\chi}_\sigma(\bm{\bm{k}})=\omega^{p}_\sigma(\bm{k}) \bm{\chi}_\sigma(\bm{k}),~~~H_{\textrm{eff}}=i\begin{pmatrix}
-A & -D(\bm{k}) \\
I_{nd} & -A
\end{pmatrix}, \label{eq.phonon_eigenvector}
\end{equation}
where $\sigma$ in the subscript is the index for eigenmodes for phonons. Here, we have combined the polarization vector for the displacement, $\bm{\epsilon} _\sigma(\bm{k})$, and momentum, $\bm{\mu}_\sigma(\bm{k})$, into a single object 
\begin{equation}
\bm{\chi}_\sigma(\bm{k})=\begin{pmatrix}
\bm{\mu}_{\sigma} (\bm{k})\\
\bm{\epsilon}_{\sigma} (\bm{k})
\end{pmatrix}  \label{eq.polarization}.
\end{equation}
Let us note that $H_{\textrm{eff}}=2\rho_{y} H_{p}$, where $\rho_{i}$ with $i=x,y,z$ are Pauli matrices in the phonon sector, defining the block structure in Eq.~\eqref{eq.phonon_eigenvector}. The lower block of the matrix equation~\eqref{eq.phonon_eigenvector} is just the relation between the conjugate momentum and the kinetic momentum: $\bm{p}_\alpha=\dot{\bm{u}}_\alpha+\Lambda_\alpha \bm{u}_\alpha$, where the dot $(\dot{~})$ is the differentiation with respect to time, implies 
\begin{equation}
\bm{\mu}_{\sigma} (\bm{k})=-i\omega^{p}_{\sigma}(\bm{k})\bm{\epsilon}_{\sigma} (\bm{k})+A\bm{\epsilon}_{\sigma} (\bm{k}). \label{eq.polarization_relation}
\end{equation} 
By using Eqs.~\eqref{eq.phonon_eigenvector}, one can show that the eigenvalues and eigenvectors of $H_{\textrm{eff}}$ always come in pairs, $\sigma,-\sigma $ with the following relation: $\bm{\chi}_{-\sigma}^*(-\bm{k})=\bm{\chi}_\sigma(\bm{k})$  and $\omega^{p}_{-\sigma}(-\bm{k})=-\omega^{p}_{\sigma}(\bm{k}) $. Here, we have used the convention where $\sigma>0$ corresponds to $\omega^{p}_{\sigma}(\bm{k}) \geq 0$. Let us note that $\sigma$ takes values in $-s d, -s d + 1 ..., -1, 1, ..., sd-1,sd$, where $d=2$ is the dimension in which the vibration takes place. 

The eigenvectors can be normalized as follows:
\begin{equation}
\bm{\chi}_{\sigma}(\bm{k})^\dagger \rho_{y} \bm{\chi}_{\sigma'}(\bm{k})=(\rho_{z})_{\sigma \sigma'}. \label{eq.chi_normalization}
\end{equation}
Although it is possible to give a direct proof of this\cite{zhang2011phonon,zhang2010topological} [see also Appendix~\ref{ap.other_formulation}], we will instead assume that this normalization condition is given, and then show in the next section that when we transform the phonon Hamiltonian to a bosonic BdG Hamiltonian, $\bm{\chi}_{\sigma}(\bm{k})$ are mapped to the eigenvectors of bosonic BdG Hamiltonian, see Eq.~\eqref{eq.phonon_bdg}, \eqref{eq.phonon_bdg_normalization}, and \eqref{eq.PhononToBdG}. 
This gives an alternative proof of the normalization condition. 
We also note that the convention we use to normalize $\bm{\chi}_{\sigma}(\bm{k})$ differs from the normalization condition in Refs.~[\onlinecite{zhang2011phonon,zhang2010topological}], where the authors use $\rho_y/2\omega_\sigma(\bm{k}) $ as the metric for $\bm{\chi}_{\sigma}(\bm{k}) $ (the precise relation is discussed in Appendix \ref{ap.other_formulation}). 
We prefer the normalization given here because it behaves well even for acoustic phonon modes and the relation between $\bm{x}(\bm{k})$ and the phonon operators $b_{\bm{k},\sigma}$ is simpler, see Eq.~\eqref{eq.phonon_second_quantization}. 
This does not affect the Berry curvature that we define in Sec.~\ref{ssec.BdG_BC}, which we show in Appendix \ref{ap.other_formulation}. 

Finally, let us note that the completeness relation is given by
\begin{equation}
\sum_\sigma \bm{\chi}_\sigma (\bm{k})  (\rho_{z})_{\sigma \sigma} \bm{\chi}_{\sigma}^\dagger (\bm{k}) \rho_{y}=I_{2sd}, \label{eq.phonon_completeness}
\end{equation}
where $I_{2sd}$ is the $2sd\times 2sd$ identity matrix. This relation  can be checked by multiplying the right hand side by $\bm{\chi}_{\sigma'}(\bm{k}) $.

\subsection{Second quantization: relation to BdG Hamiltonian} \label{ssec.BdG}

We give a matrix formulation of the second quantization problem of phonon in the presence of magnetic field. It will be shown that this is only a simple variation of the BdG problem. Then, we use this to show how the hybridization problem of magnon and phonon  can be mapped to the BdG problem. This relation gives us a simple method to diagonalize the magnetoelastic Hamiltonian. In the next subsection, we use this diagonalization method to define the Berry connection.

In order to understand the relation between the phonon problem and the BdG Hamiltonian, let us first note that the metric used on the normalization of the polarization vectors $\bm{\chi}_\sigma(\bm{k})$ also appears in the commutation relation between the operators, $[\bm{x}_{\sigma}(\bm{k})^\dagger,\bm{x}_{\sigma'}(\bm{k}')]=(\rho_y)_{\sigma \sigma'}\delta_{\bm{k}\bm{k'}}$. \footnote{If we restore $\hbar$, we can absorb the $\hbar$ into the Hamiltonian by redefining $\bm{x}\rightarrow \bm{x}/\sqrt{\hbar}$ and $H_{{\textrm{eff}}}\rightarrow \hbar H_{\textrm{eff}}$.}
Next, we note that the field operators $\bm{y}_{\bm{k}}$ in bosonic BdG Hamiltonian satisfy 
\begin{align}
(i)&~ [\bm{y}^{\dagger}_{\sigma}(\bm{k}) ,\bm{y}_{\sigma'}(\bm{k}')  ]=-(\rho_{z})_{\sigma \sigma'}\delta_{\bm{k}\bm{k}'} \label{cond_i}\\
(ii)&~ \bm{y}_{\sigma }^{\dagger}(\bm{k}) =(\rho_{x})_{\sigma \sigma'}\bm{y}_{\sigma'}(-\bm{k}) . \label{eq.cond_ii}
\end{align}
The condition (i) can be satisfied by making use of the following:
\begin{equation}
U^\dagger(\theta) \rho_{y} U(\theta)=\cos 2\theta \rho_{y}+\sin 2\theta \rho_{z},~U(\theta)=e^{i \rho_{x} \theta}.
\end{equation}
For $\theta=\pi/4$,  $U(\frac{\pi}{4})=\frac{\sqrt{2}}{2}(1+i  \rho_{x})$ and $\rho_{y}\rightarrow \rho_{z}$.
Therefore, if we define $\bm{\chi}_\sigma(\bm{k}) =U(\frac{\pi}{4})\tilde{\bm{\xi}}_\sigma (\bm{k}) $, the normalization condition is 
\begin{equation}
\tilde{\bm{\xi}}_{\sigma}^\dagger (\bm{k})\rho_{z} \tilde{\bm{\xi}}_{\sigma'} (\bm{k})=(\rho_{z})_{\sigma \sigma'}.
\end{equation}
Similarly, if we define $\bm{x}(\bm{k}) =U(\frac{\pi}{4})\tilde{\bm{y}}(\bm{k}) $
\begin{align}
[\tilde{\bm{y}}^\dagger_{\sigma}(\bm{k}),\tilde{\bm{y}}_{\sigma'}(\bm{k})]=-(\rho_{z})_{\sigma \sigma'}.
\end{align}

To satisfy the condition (ii), let us note that we can make an additional transformation that fixes the metric $\rho_{z}$. Let us define
\begin{equation}
U'=\begin{pmatrix}
I_{sd} & \\
 & -i I_{sd}
\end{pmatrix}.
\end{equation}
If we define $\tilde{\bm{\xi}}_\sigma (\bm{k}) =U' \bm{\xi}_\sigma (\bm{k})$ and $\tilde{\bm{y}}(\bm{k}) =U' \bm{y}(\bm{k}) $, we can write 
\begin{equation}
\bm{\chi}_{\sigma}(\bm{k})=V\bm{\xi}_{\sigma}(\bm{k}), \quad
\bm{x}(\bm{k})=V\bm{y}(\bm{k}) , \label{eq.phonon_bdg}
\end{equation}
where 
\begin{equation}
V=U\left(\frac{\pi}{4}\right)U' ,\label{eq.V_matrix}
\end{equation}
and 
\begin{align}
\bm{\xi}_{\sigma}^\dagger (\bm{k})\rho_{z} \bm{\xi}_{\sigma'} (\bm{k})&=(\rho_{z})_{\sigma \sigma'} \label{eq.phonon_bdg_normalization}\\
 [\bm{y}^\dagger_{\sigma}(\bm{k}),\bm{y}_{\sigma'}(\bm{k})]&=-(\rho_{z})_{\sigma \sigma'}.
\end{align}
Moreover, we have
\begin{equation}
\bm{y}(\bm{k}) =\frac{\sqrt{2}}{2}\begin{pmatrix}
\bm{p}(\bm{k})-i \bm{u}(\bm{k}) \equiv\bm{v}_{\bm{k}}\\
\bm{p}(\bm{k})+i \bm{u}(\bm{k})= \bm{v}^\dagger_{-\bm{k}}
\end{pmatrix}. \label{eq.phonon_BdGbasis}
\end{equation}
so that condition (ii) is satisfied.

With the transformation discussed above, the eigenvalue problem for $\bm{\xi}_\sigma(\bm{k})$ becomes
\begin{align}
[V^\dagger \rho_{y} H_{p}(\bm{k})  V]\bm{\xi}_\sigma(\bm{k})&=[\rho_{z} V^\dagger H_{p}(\bm{k})  V]\bm{\xi}_\sigma(\bm{k})\nonumber \\
&=\frac{\omega^{p}_{\sigma}(\bm{k})}{2} \bm{\xi}_\sigma(\bm{k}),
\label{eq.PhononToBdG}
\end{align}
where the eigenvectors satisfy the constraint given in Eq.~\eqref{eq.phonon_bdg_normalization}
In addition, because the field operator $\bm{y}(\bm{k})$ is a bosonic BdG field,  $V^\dagger H_{p}(\bm{k}) V$ is a bosonic BdG Hamiltonian, and we see that the constraint given in Eq.~\eqref{eq.phonon_bdg_normalization} is just the constraint on the eigenvectors of a bosonic BdG Hamiltonian. At $\bm{k}$ points where $H_p(\bm{k})$ has no zero modes, it is  positive definite, and $\bm{\xi}_{\sigma}$ satisfying Eqs.~\eqref{eq.phonon_bdg_normalization} and \eqref{eq.PhononToBdG} can be found using the Colpa's method. This gives an alternative justification of the normalization condition in Eq.~\eqref{eq.chi_normalization}.

To write the Hamiltonian in terms of the phonon operators, let us make the expansion \begin{equation}
\bm{x}(\bm{k})=\sum_{\sigma}\bm{\chi}_{\sigma}(\bm{k})b_{\bm{k},\sigma} \label{eq.phonon_second_quantization}
\end{equation}
or equivalently,  $\bm{u}(\bm{k})=\sum_{\sigma}\bm{\epsilon}_{\sigma}(\bm{k})b_{\bm{k},\sigma}$ and  $\bm{p}(\bm{k})=\sum_{\sigma}\bm{\mu}_{\sigma}(\bm{k})b_{\bm{k},\sigma}$, where $b_{\bm{k},\sigma}$ satisfies the canonical commutation relation for $\sigma>0$ and $b_{\bm{k},\sigma}^\dagger=b_{-\bm{k},-\sigma}$, so that $[b_{\bm{k},\sigma},b_{\bm{k},\sigma'}^\dagger]=\delta_{\bm{k},\bm{k}'}(\rho_z)_{\sigma \sigma'}$. Then,  we have
\begin{align}
{\cal H}_p&=\frac{1}{2}\sum_{\bm{k}}
\bm{x}(-\bm{k})^{T} \rho_{y} H_{\textrm{eff}}(\bm{k}) \bm{x}(\bm{k})
 \nonumber \\
&=\frac{1}{2}\sum_{\bm{k},\sigma,\sigma'}
\bm{\chi}_{\sigma'}^{T}(-\bm{k})\rho_{y} H_{\textrm{eff}}(\bm{k}) 
\bm{\chi}_{\sigma}(\bm{k})
b_{-\bm{k},\sigma'} b_{\bm{k},\sigma}\nonumber \\
&=\frac{1}{2}\sum_{\bm{k},\sigma,\sigma'}
\bm{\chi}_{\sigma'}^\dagger(\bm{k})
\rho_{y} \omega^p_{\sigma} (\bm{k})\bm{\chi}_{\sigma}(\bm{k}) b_{-\bm{k},-\sigma'} b_{\bm{k},\sigma}\nonumber \\
&=\frac{1}{2}\sum_{\bm{k},\sigma}|\omega^{p}_{\sigma}(\bm{k}) | b_{-\bm{k},-\sigma} b_{\bm{k},\sigma}\nonumber \\
&=\frac{1}{2}\sum_{\bm{k},\sigma}|\omega^{p}_{\sigma}(\bm{k}) |b^\dagger_{\bm{k},\sigma} b_{\bm{k},\sigma}.
\end{align} 
In the third line, we used the identity $\bm{\chi}_\sigma(\bm{k})^*=\bm{\chi}_{-\sigma}(-\bm{k})$.

\subsection{Diagonalization of Magnetoelastic Hamiltonian} \label{ssec.diag_me}

We will now develop a simple method to diagonalize the magnetoelastic Hamiltonian $H_{me}$ defined in Eq.~\eqref{eq.magnetoelastic_hamiltonian}. 
To do this, we first keep track of the matrices that are used to introduce phonon operators and diagonalize the resulting bosonic BdG Hamiltonian. 
Then, we will observe that if we introduce $H_{s}$ defined in Eq.~\eqref{eq.simplified_hamiltonian}, the diagonalization procedure can be simplified. 
Because this will require us to introduce various forms of Pauli matrices, let us first explain the notation that will be used.

Let us define $\rho_i$ to be the $2sd\times 2sd$ Pauli matrices for the $\bm{x}$ and $\bm{p}$ blocks in the phonon sector. Similarly, let us define  $\tau_{i}$ to be the $2m \times 2m$ Pauli matrices for the particle and hole blocks in the magnon sector. Here, $m$ is the number of HP operators in a unit cell, and we assume that the magnon Hamiltonian is written in BdG form. When there is no source for confusion, we will abuse the notation and write $\rho_i$ to mean $I_{2m}\oplus \rho_i$ and $\tau_i$ to mean $\tau_i \oplus I_{2sd}$. 
In the same spirit, it is to be understood that
\begin{equation}
V=I_{2m}\oplus V,
\end{equation}
where $V$ on the right hand side was defined in Eq.~ \eqref{eq.V_matrix}. Finally, we will use $\sigma_{i}$ for the $2(m+sd) \times 2(m+sd)$ Pauli matrices in the magnetoelastic sector.

Let $\Phi_{\bm{k}}=(a_{\bm{k},1}, ... , a_{-\bm{k},1}^\dagger, ..., \bm{p}^T_1(\bm{k}),..., \bm{u}^T_1(\bm{k}),...)$, where $a_{i,\bm{k}}$ for $i=1,...,m$ are the HP operators.
The magnetoelastic Hamiltonian is 
\begin{equation}
{\cal H}_{me}=\sum_{\bm{k}}\Phi_{\bm{k}}^{\dagger} H_{me}(\bm{k}) \Phi_{\bm{k}},
\end{equation}
where
\begin{equation}
H_{me}(\bm{k}) =\begin{pmatrix}
H_{m}(\bm{k})  & H_{c}(\bm{k})  \\
H_{c}^\dagger (\bm{k})  & H_{p}(\bm{k}) 
\end{pmatrix}, \label{eq.magnetoelastic_hamiltonian}
\end{equation}
$H_{m} (\bm{k}) $ is the magnon Hamiltonian written in BdG form [for a simple example, see Eq.~\eqref{eq.simple_magnon_example}], and $H_{c} (\bm{k}) $ is the magnon-phonon coupling Hamiltonian.

Let us now keep track of the matrices that are used to diagonalize $H_{me}$.
The transformation of phonon Hamiltonian from $\bm{u}(\bm{k})$, $\bm{p}(\bm{k})$ basis to the phonon basis is $\Phi_{\bm{k}}^{\dagger} H_{me}(\bm{k}) \Phi_{\bm{k}}=\Psi_{\bm{k}}^{'\dagger } X_{{\bm{k}} }^{\dagger} H_{me}(\bm{k})  X_{\bm{k}}\Psi'_{\bm{k}}$,  where
\begin{equation}
X_{\bm{k}}=\begin{pmatrix}
I_{2m} & 0 & ... & 0 & ...\\
0& \bm{\chi}_{1}(\bm{k}) & ...& \bm{\chi}_{-1}(\bm{k}) & ...
\end{pmatrix}
\end{equation}
and  $\Psi'_{\bm{k}}=(a_{1\bm{k}}, ...,  a^{\dagger}_{1-\bm{k}}, ...,  b_{\bm{k},1}, ... , b^{\dagger}_{\bm{k},-1},...)$. Note that $\Psi'_{\bm{k}}$ is the field operators that is usually used to write the magnon-phonon Hamiltonian \cite{takahashi2016berry,flebus2017magnon,thingstad2018chiral}. 
Note that we have
\begin{equation}
X_{\bm{k}}^\dagger \tau_{z}\rho_{y} X_{\bm{k}}=\tau_{z} \rho_{z} \label{eq.X_normalization},
\end{equation}
where we have used Eq.~\eqref{eq.chi_normalization}. Let $P$ be the permutation that swaps half the magnon sector with half the phonon sector:
\begin{equation}
P=\begin{pmatrix}
I_{m} & & & \\
& & I_{sd} & \\
& I_{m} & & \\
& & & I_{sd}
\end{pmatrix} \label{eq.permutation_matrix}.
\end{equation}
so that $\Psi_{\bm{k}}\equiv P\Psi'_{\bm{k}}=(a_{1\bm{k}}, ...,  b_{\bm{k},1}, ..., a^{\dagger}_{1-\bm{k}} , ... , b^{\dagger}_{\bm{k},-1},...)$. Then, 
\begin{equation}
\tilde{H}_{me}(\bm{k}) \equiv P X^{\dagger}_{\bm{k}} H_{me}(\bm{k})  X_{\bm{k}} P^\dagger \label{eq.magnon_phonon_hamiltonian}
\end{equation}
and
\begin{equation}
{\cal H}_{me}=\sum_{\bm{k}}\Psi_{\bm{k}}^{\dagger} \tilde{H}_{me}(\bm{k}) \Psi_{\bm{k}},
\end{equation}
is the magnetoelastic Hamiltonian in HP operator and phonon operator basis arranged in BdG form. 
Let $T_{mp}(\bm{k}) $ be the transformation that diagonalizes  $\tilde{H}_{me}(\bm{k}) $ satisfying the normalization condition
\begin{equation}
T_{mp}(\bm{k})^\dagger \sigma_z T_{mp}(\bm{k}) =\sigma_z .\label{eq.T_mp_normalization}
\end{equation}
Then, $X_{\bm{k}} P^\dagger T_{mp}(\bm{k}) $ is the transformation that diagonalizes $H_{me}(\bm{k}) $, satisfying the normalization condition 
\begin{equation}
[ T_{mp}^{\dagger}(\bm{k})  P X_{\bm{k}}^\dagger]\tau_{z} \rho_{y}  [X_{\bm{k}} P^\dagger T_{mp}(\bm{k}) ]=\sigma_{z},
\end{equation}
where we have used Eq.~\eqref{eq.X_normalization}, $P \tau_z \rho_zP^\dagger=\sigma_z$, and Eq.~\eqref{eq.T_mp_normalization}.

Now, let us simplify the diagonalization process. First, define the `simplified' Hamiltonian as
\begin{equation}
H_s(\bm{k}) =P V^\dagger H_{me}(\bm{k})  V P^\dagger .\label{eq.simplified_hamiltonian}
\end{equation}
Then, the matrix that diagonalizes the Hamiltonian is given by
\begin{equation}
T_{s}(\bm{k}) =P V^\dagger X_{\bm{k}} P^\dagger T_{mp}(\bm{k}).\label{eq.simplified_eigenvector}
\end{equation}
In other words,
\begin{equation}
T_{s}^\dagger (\bm{k}) H_s (\bm{k}) T_s(\bm{k}) =\textrm{diag}(|\omega^{mp}_{n \bm{k}}|/2) \label{eq.H_S_diagonalization}
\end{equation}
and
\begin{equation}
T_s^\dagger(\bm{k}) \sigma_{z} T_s(\bm{k})=\sigma_{z} \label{eq.T_s}
\end{equation}
Now, the problem of finding $T_{s}(\bm{k})$ that satisfies Eqs.~\eqref{eq.H_S_diagonalization} and \eqref{eq.T_s} is exactly that of solving a bosonic BdG Hamiltonian. 
Thus, the $H_s$ can be diagonalized using Colpa's method \cite{colpa1978diagonalization}. 
For our purposes, we may assume that the matrix that diagonalizes $H_s(\bm{k})$ obtained from Colpa's method is given by Eq.~\eqref{eq.simplified_eigenvector}. This is because the column vectors of $T_{s}(\bm{k})$ is unique up to a phase factor when there is no degeneracy in the eigenvalues of $H_{s}(\bm{k})$ [see Appendix~\ref{ap.BdG_review}].
\textit{In conclusion, if we start with $H_s$, we can just use Colpa's method to diagonalize the Hamiltonian to solve the magnon-phonon problem. }

Finally, let us note that there is a simple explanation for the reason that the $H_{s}(\bm{k})$ can be diagonalized by using the Colpa's method.
Let us write
\begin{align}
\Phi_{\bm{k}}^\dagger H_{me}(\bm{k}) \Phi_{\bm{k}}=\Sigma_{\bm{k}}^\dagger H_{s}(\bm{k}) \Sigma_{\bm{k}}
\end{align}
where we have defined
\begin{align}
\Sigma_{\bm{k}}&\equiv PV^\dagger \Phi_{\bm{k}} \nonumber \\
&=P\begin{pmatrix}
a_{\bm{k},1} \\
\vdots\\
a_{-\bm{k},1}^\dagger \\
\vdots\\
\bm{v}_{\bm{k}} \\
\bm{v}_{-\bm{k}}^\dagger
\end{pmatrix}
=\begin{pmatrix}
a_{\bm{k},1} \\
\vdots \\
\bm{v}_{\bm{k}} \\
a_{-\bm{k},1}^\dagger \\
\vdots\\
\bm{v}_{-\bm{k}}^\dagger
\end{pmatrix}
\end{align}
where $\bm{v}_{\bm{k}}$ is the phonon BdG field which was defined in Eq.~\eqref{eq.phonon_BdGbasis}. 
Because $\Sigma_{\bm{k}}$ is bosonic BdG field, $H_{s}(\bm{k})$ is bosonic BdG Hamiltonian, and therefore can be solved using the Colpa's method.
The reason we have given a complicated derivation is to obtain Eq.~\eqref{eq.simplified_eigenvector}, which relates $T_{mp}(\bm{k})$ and $T_{s}(\bm{k})$

\section{Berry connection}
\label{sec.berry}
In this section, we first briefly review the Berry connection for bosonic BdG Hamiltonian. Then, we define the Berry connection for the simplified magnetoelastic Hamiltonian $H_{s}(\bm{k}) $, and show that it is different from that defined using $\tilde{H}_{me}(\bm{k}) $ in which the phonon part is already diagonalized.  This will have physical consequences because the Berry curvature is related to the anomalous velocity of a semiclassical wave packet.

\subsection{Berry connection of BdG systems}
\label{ssec.BdG_BC}

Let us first review some useful properties of a positive definite bosonic BdG Hamiltonian.
These properties apply to phonon, magnon, and magnetoelastic Hamiltonians. 
Let $H_{\bm{k}}$ be a $2N\times 2N$  BdG Hamiltonian.
The matrix $T_{\bm{k}}$ that diagonalizes the BdG Hamiltonian, i.e. $T_{\bm{k}}^\dagger H_{\bm{k}} T_{\bm{k}} =\frac{1}{2}\tilde{\omega}_{\bm{k}}$,  satisfy $T_{\bm{k}}^\dagger \sigma_z T_{\bm{k}}=\sigma_z$ because different choices of field operators should preserve the bosonic commutation relation. 
We note that the diagonal matrix $\tilde{\omega}_{\bm{k}}$ has positive (diagonal) components.
We also have 
\begin{equation}
H_{-\bm{k}}=\sigma_{x}H^*_{\bm{k}} \sigma_{x}, \quad T_{-\bm{k}}=\sigma_x T^*_{\bm{k}}\sigma_x \label{eq.bdg_negk}
\end{equation}
because of the condition (ii) satisfied by the BdG field operators [see Eq.~\eqref{eq.cond_ii}]. The eigenvalues $\frac{\omega_{\bm{k},n}}{2}$ and eigenvectors $|T_{\bm{k}}\rangle_{n}$, which are column vectors of the matrix $T_{\bm{k}}$, satisfy $\sigma_{z} H_{\bm{k}}|T_{\bm{k} } \rangle_{n} = \frac{\omega_{\bm{k},n} }{2} | T_{\bm{k}} \rangle_{n}$. 
Note that if we define $\omega_{\bm{k}}=\textrm{diag}(\omega_{\bm{k},n})$, the eigenvalue problem is equivalent to $\sigma_z H_{\bm{k}}T_{\bm{k}}=\frac{1}{2}T_{\bm{k}}\omega(\bm{k})$, or  $T^\dagger_{\bm{k}} H_{\bm{k}}T_{\bm{k}}=\frac{1}{2}\sigma_z \omega(\bm{k})\equiv \frac{1}{2}\tilde{\omega}_{\bm{k}}$.
Using Eq.~\eqref{eq.bdg_negk}, it can be shown that the eigenvalues and eigenvectors come in pairs: If we let $n>0$ correspond to $\omega_{\bm{k},n}> 0$, $\omega_{\bm{k},-n}=-\omega_{-\bm{k},n}$ and $|T_{\bm{k}}\rangle_{-n}=\sigma_{x}|T_{-\bm{k}}^{*}\rangle_{n}$. 

Next, let us review the Berry connection of a bosonic BdG Hamiltonian. 
We first note that the gauge group for $H_{\bm{k}}$ is the indefinite unitary group $U(N,N)$ whose elements $G(\bm{k})$  satisfy $G^\dagger(\bm{k}) \sigma_z G(\bm{k})=\sigma_z$ (for sub-bands, it is a subgroup of $U(N,N)$). 
This is because different choices of field operators should preserve the bosonic commutation relation.
For convenience, we define the quantity
\begin{equation}
\bm{{\cal \bm{A}}}_{n n'}(\bm{k})=  \prescript{}{n}\langle T_{\bm{k}}|i\sigma_z \bm{\nabla} |T_{\bm{k}}\rangle _{n'}.\label{eq.non-abelian_BC}
\end{equation}
The non-Abelian Berry connection is defined by \cite{nakahara2003geometry,shindou2013topological} 
\begin{equation}
\bm{A}_{nn'}(\bm{k})=(\sigma_{z} \bm{{\cal \bm{A}}}(\bm{k}))_{nn'}.
\end{equation}
Under the gauge transformation $|T_{\bm{k}}\rangle_{n} \rightarrow |T_{\bm{k}}\rangle_{n'} G_{n'n}(\bm{k})$, where $G(\bm{k})\in U(N,N)$, 
\begin{equation}
\bm{{\cal A}}(\bm{k}) \rightarrow G^\dagger(\bm{k})\bm{{\cal A}}(\bm{k}) G(\bm{k})+G^\dagger(\bm{k}) i \sigma_z \bm{\nabla} G(\bm{k})
\end{equation}
so that 
\begin{equation}
\bm{A}(\bm{k}) \rightarrow G^{-1}(\bm{k})\bm{A}(\bm{k}) G(\bm{k})+G^{-1} (\bm{k})i\bm{\nabla} G(\bm{k}),
\end{equation}
since $G^{-1}(\bm{k})=\sigma_z G(\bm{k})^\dagger \sigma_z$.

Thus, the Berry curvature is given by
\begin{equation}
\bm{B}_n (\bm{k})= \bm{\nabla} \times \bm{A}_{nn}(\bm{k}) \label{eq.berry_curvature}
\end{equation}
and the Chern number is given by
\begin{equation}
C_n=\frac{1}{2\pi} \int_{BZ} d\bm{k}B^z_n(\bm{k}).
\end{equation}
The sum rule for Chern number \cite{shindou2013topological} states that the total Chern number for the sector $n>0$ is zero. This allows us to define the Chern number when the lowest energy band has zero modes because the total Chern number vanishes regardless of how we open up the gap. It can also be shown  that $\bm{B}_{-n}(\bm{k})=-\bm{B}_{n}(-\bm{k})$ so that $C_{n}=-C_{-n}$ \cite{matsumoto2014thermal}.

\subsection{Magnon-Polaron Berry connection}
\label{ssec.magnetoelastic_BC}
Let us derive the difference between the Berry connection defined by using $H_s$ and $\tilde{H}_{me}$. For notational simplicity, let us work with $\bm{{\cal A}}(\bm{k})$ instead of $\bm{A}(\bm{k})=\sigma_z \bm{{\cal A}}(\bm{k})$, and simply refer to $\bm{{\cal A}}(\bm{k})$ as the Berry connection in this subsection for magnon, phonon, and magnon-polaron. From Eq.~\eqref{eq.simplified_eigenvector} and \eqref{eq.non-abelian_BC}, we have
\begin{align}
\bm{{\cal A}}_{n n'}^{s}(\bm{k})=&\prescript{}{n}\langle T_{mp}^{\dagger}(\bm{k}) P X_{\bm{k}}^\dagger V P^\dagger  | i\sigma_{z}\bm{\nabla}|P V^\dagger X_{\bm{k}}  P^{\dagger} T_{mp}(\bm{k})\rangle_{n'} \nonumber \\
=&\prescript{}{n}\langle T_{mp}^{\dagger} (\bm{k}) P X_{\bm{k}}^\dagger | i \tau_{z} \rho_{y}\bm{\nabla}|X_{\bm{k}}  P^{\dagger} T_{mp}(\bm{k})\rangle_{n'}  \nonumber \\
=&\prescript{}{n}\langle T_{mp}^\dagger(\bm{k}) P  |
\begin{pmatrix}
0 & \\
& \bm{{\cal \bm{A}}}^p(\bm{k})
\end{pmatrix} 
|P^{\dagger}T_{mp}(\bm{k})\rangle_{n'} \nonumber \\
&+\prescript{}{n}\langle T_{mp}(\bm{k})| PX_{\bm{k}}^{
\dagger}i\tau_z \rho_y X_{\bm{k}}P^\dagger \bm{\nabla} | T_{mp} (\bm{k})\rangle_{n'} \nonumber \\
=&\prescript{}{n}\langle T_{mp}^\dagger(\bm{k}) P  |
\begin{pmatrix}
0 & \\
& \bm{{\cal \bm{A}}}^p(\bm{k})
\end{pmatrix} 
|P^{\dagger}T_{mp}(\bm{k})\rangle_{n'} \nonumber \\
&+\bm{{\cal A}}^{mp}_{n n'}(\bm{k}),
\end{align}
where the phonon Berry connection is 
\begin{equation}
\bm{{\cal A}}^P_{\sigma \sigma'}(\bm{k})=\langle \bm{\chi}_\sigma (\bm{k})|i \rho_y \bm{\nabla|} \bm{\chi}_{\sigma'} (\bm{k})\rangle=\langle \bm{\xi}_\sigma(\bm{k}) |i \rho_z \bm{\nabla|} \bm{\xi}_{\sigma'}(\bm{k}) \rangle
\end{equation}
and 
\begin{equation}
\bm{{\cal A}}^{mp}(\bm{k})=\prescript{}{n}\langle T_{mp}(\bm{k})| i\sigma_z \bm{\nabla} | T_{mp} (\bm{k})\rangle_{n'}.
\end{equation}

We should note that the contribution from the phonon (the first term in the above equation) does not vanish in general. To understand this, let us define $W_{\bm{k}}=PV^{\dagger}X_{\bm{k}} P^{\dagger}$. Since we are interested in the Berry curvature of individual bands, the only allowed transformation is of the form $|T_{s}(\bm{k})\rangle_n \rightarrow |T_{s}(\bm{k})\rangle_n e^{i\zeta_{n}(\bm{k})}$ which does not mix different bands. However, $|T_{mp} (\bm{k}) \rangle \rightarrow W_{\bm{k}} |T_{mp} (\bm{k})\rangle$ is different in nature. Even though it does not change the energy, it mixes the matrix components of $|T_{mp} (\bm{k}) \rangle$,  so that one can expect a change in the Berry curvature. For an explicit comparison, see  Fig.~\ref{fig.mp_berry}.

\section{Conclusions} \label{sec.conclusion}

We have explained how to compute the Berry curvature and Chern number in magnon-polaron bands by finding the relation between magnetoelastic Hamiltonian and the bosonic BdG Hamiltonian. 
As an example, we have applied this to the triangular antiferromagnet with out-of-plane magnetic field, where the magnon and phonon bands have zero Chern number. Although the magnon-phonon coupling arising from exchange magnetostriction of Heisenberg model does not by itself generate Berry curvature, it induces, in the presence of out-of-plane magnetic field, large Berry curvature in the anti-crossing regions. In addition, all of the resulting magnon-polaron bands are gapped, and they carry non-zero Chern numbers. The Berry curvature arising from magnon-phonon hybridization can significantly renormalize the phonon thermal Hall conductivity. 

Similarly, we expect that magnon-phonon coupling can significantly renormalize magnon Hall conductivity. 
Recently, Ref.~[\onlinecite{kim2018magnon}] discussed possible thermal Hall conductivity in trimerized triangular lattice antiferromagnet $\textrm{YMnO}_3$. 
As the authors point out, magnon Hall conductivity may be renormalized by magnon-phonon coupling.  
Although the magnon-phonon coupling from exchange magnetostriction does not by itself induce Berry curvature in our toy model, this may not be true when there is trimerization.
Therefore, in the presence of magnon-phonon coupling in trimerized triangular antiferromagnet, we can expect significant deviation of thermal Hall conductivity from that resulting only from magnons. 
In practice, however, it may be difficult to distinguish the contribution to Hall response from magnon-phonon coupling and uncertainty in parameters in magnetoelastic Hamiltonian. 
Therefore, an interesting question would be to ask whether it is possible for the thermal Hall conductivity or spin Nernst conductivity to arise solely from magnon-phonon coupling.  We leave these questions for future research. 

\section{Acknowlegmenet}
We thank Je-Geun Park and Joosung Oh for their kind explanation of the neutron scattering data published in Ref.~[\onlinecite{oh2016spontaneous}], which motivated us to work on this project.
S.P. was supported by IBS-R009-D1.
B.-J.Y. was supported by the Institute for Basic Science in Korea (Grant No. IBS-R009-D1) and Basic Science Research Program through the National Research Foundation of Korea (NRF) (Grant No. 0426-20170012, No.0426-20180011), and  the POSCO Science Fellowship of POSCO TJ Park Foundation (No.0426-20180002).
This work was supported in part by the U.S. Army Research Office under Grant Number W911NF-18-1-0137.

\appendix

\section{Symmetry Analysis} \label{ap.symmetry}

In this section, we discuss the symmetry representation of magnon, phonon, and magnon-polaron. For clarity, we put a hat ($\hat{~}$) over operators in this section.

\subsection{Magnon Symmetry} \label{ap.magnon_symmetry}

In this subsection, we will study the symmetry of Heisenberg Hamiltonian. 
Let us first assume that there is no external magnetic field, i.e. no canting. 
Because the Heisenberg model arises when there is no spin-orbit coupling, the (unitary) symmetry of Heisenberg model on triangular lattice as a wallpaper group is $p6mm \otimes SU(2)$ when there is no magnetization. 
When there is magnetic ordering, the symmetry will be lowered to a subgroup of this symmetry group.
In particular, the symmetry of the ground state is generated by the following symmetry operators (the right hand side defines the action on the spin position and direction, respectively):
\begin{align}
\hat{{\cal T}}_{\bm{R}_1}&=T_{\bm{R}_1} \otimes C_3 \nonumber \\
\hat{\cal T}_{\bm{R}_2}&= T_{\bm{R}_2} \otimes C_3^{-1} \nonumber \\
\hat{\cal C}_{3z} &= C_{3z} \otimes C_{3z} \nonumber \\
\hat{\cal{C}}_{2x} &= C_{2x} \otimes 1 \nonumber \\
\hat{\cal{C}}_{2y} &=C_{2y} \otimes C_{2y} \label{eq.gs_symmetry}
\end{align}
Here, $T_{\bm{R}_i}\otimes 1$ is a lattice translation by $\bm{R}_{i}$, $C_{3z}\otimes 1$ is a rotation of the lattice positions by $120^\circ$ counterclockwise about the center of yellow triangle, $C_{2x}\otimes 1$ is a twofold rotation about the line through the lattice sites $1$ and $2$, $C_{2y} \otimes 1$ is a twofold rotation about the line through lattice sites $4$ and $3$, and $1\otimes( C_{2x},C_{2y},C_{3z})$ are vectorial rotation of spin directions.

\begin{figure}[t!]
\centering
\includegraphics[width=8cm]{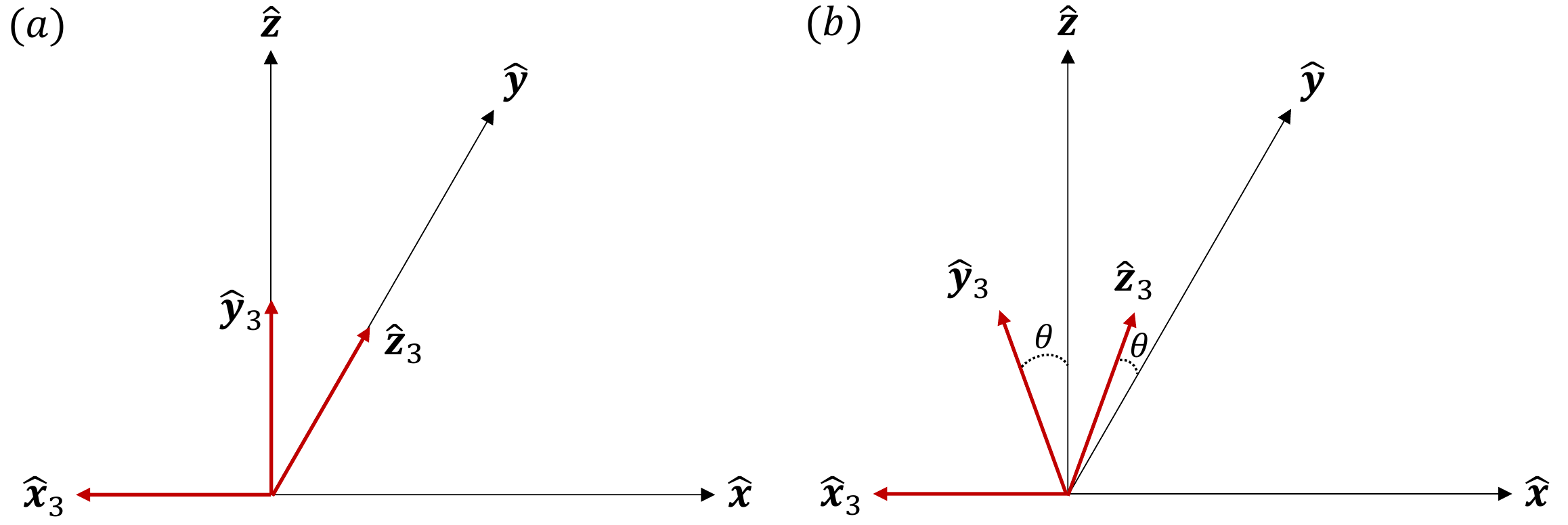}
\caption{Global axes ($\hat{\bm{x}},\hat{\bm{y}},\hat{\bm{z}}$) and local axes of spin at site $3$ in Fig.~\ref{fig.triangular_lattice} ($\hat{\bm{x}}_3,\hat{\bm{y}}_3,\hat{\bm{z}}_3$) (a) without canting, (b) with canting by angle $\theta$. The local axes for sites $1$ ($2$) are obtained through rotation by $120^\circ$ about the global $\hat{\bm{z}}$ axis counterclockwise (clockwise).}
\label{fig.axes}
\end{figure}

In order to introduce the HP operators, we have defined local axes as described in Fig.~\ref{fig.axes}, so that for spin at $\bm{R}_i$,
\begin{equation}
\hat{\bm{S}}_{\bm{R}_i}=\hat{S}_{\bm{R}_i,x}\hat{\bm{x}}_{\bm{R}_i}+\hat{S}_{\bm{R}_i,y}\hat{\bm{y}}_{\bm{R}_i}+\hat{S}_{\bm{R}_i,z}\hat{\bm{z}}_{\bm{R}_i}
\end{equation}
The local axes should be thought of as operators transforming under the symmetry representations in Eq.~\eqref{eq.gs_symmetry},
\begin{align}
\hat{{\cal T}}_{\bm{R}_1} (\hat{\bm{x}}_{\bm{R}_i},\hat{\bm{y}}_{\bm{R}_i},\hat{\bm{z}}_{\bm{R}_i})  \hat{{\cal T}}_{\bm{R}_1}^{-1}&= (\hat{\bm{x}}_{\bm{R}_i+\bm{R}_1},\hat{\bm{y}}_{\bm{R}_i+\bm{R}_1},\hat{\bm{z}}_{\bm{R}_i+\bm{R}_1})\nonumber \\
\hat{\cal T}_{\bm{R}_2}(\hat{\bm{x}}_{\bm{R}_i},\hat{\bm{y}}_{\bm{R}_i},\hat{\bm{z}}_{\bm{R}_i}) \hat{\cal T}_{\bm{R}_2}^{-1}&= (\hat{\bm{x}}_{\bm{R}_i+\bm{R}_1},\hat{\bm{y}}_{\bm{R}_i+\bm{R}_1},\hat{\bm{z}}_{\bm{R}_i+\bm{R}_1}) \nonumber \\
\hat{{\cal C}}_{3z}(\hat{\bm{x}}_{\bm{R}_i},\hat{\bm{y}}_{\bm{R}_i},\hat{\bm{z}}_{\bm{R}_i}) \hat{{\cal C}}_{3z}^{-1}&= (\hat{\bm{x}}_{C_{3z}\bm{R}_i},\hat{\bm{y}}_{C_{3z}\bm{R}_i},\hat{\bm{z}}_{C_{3z}\bm{R}_i}) \nonumber \\
\hat{{\cal C}}_{2x}(\hat{\bm{x}}_{\bm{R}_i},\hat{\bm{y}}_{\bm{R}_i},\hat{\bm{z}}_{\bm{R}_i}) \hat{{\cal C}}_{2x}^{-1}&= (\hat{\bm{x}}_{C_{2x}\bm{R}_i},\hat{\bm{y}}_{C_{2x}\bm{R}_i},\hat{\bm{z}}_{C_{2x}\bm{R}_i})  \nonumber \\
\hat{{\cal C}}_{2y}(\hat{\bm{x}}_{\bm{R}_i},\hat{\bm{y}}_{\bm{R}_i},\hat{\bm{z}}_{\bm{R}_i}) \hat{{\cal C}}_{2y}^{-1}&= (-\hat{\bm{x}}_{C_{2y}\bm{R}_i},-\hat{\bm{y}}_{C_{2y}\bm{R}_i},\hat{\bm{z}}_{C_{2y}\bm{R}_i}) \nonumber \\
\hat{{\cal C}}'_{2y}(\hat{\bm{x}}_{\bm{R}_i},\hat{\bm{y}}_{\bm{R}_i},\hat{\bm{z}}_{\bm{R}_i}) \hat{{\cal C}}_{2y}^{'-1}&= (\hat{\bm{x}}_{C_{2y}\bm{R}_i},\hat{\bm{y}}_{C_{2y}\bm{R}_i},\hat{\bm{z}}_{C_{2y}\bm{R}_i}), \label{eq.local_axes_transformation}
\end{align}
and
\begin{align}
\hat{{\cal C}} (\hat{S}_{\bm{R}_i,x},\hat{S}_{\bm{R}_i,y},\hat{S}_{\bm{R}_i,z})  \hat{{\cal C}}^{-1}&= (\hat{S}_{C\bm{R}_i,x},\hat{S}_{C\bm{R}_i,y},\hat{S}_{C\bm{R}_i,z}), \label{eq.local_spin_transformation}
\end{align}
where $\hat{{\cal C}}$ is any of the operators in Eq.~\eqref{eq.local_axes_transformation} and $C$ is its action on lattice position.
In Eq.~\eqref{eq.local_axes_transformation} we have defined an additional symmetry operator $\hat{{\cal C}}_{2y}'$, which is similar to the two-fold rotation $\hat{{\cal C}}_{2y}$. However, unlike $\hat{C}_{2y}$, which introduces a negative sign for local $x$ and $y$ axes, $\hat{{\cal C}}_{2y}'$ only changes the position indices of the local axes. $\hat{{\cal C}}_{2y}'$ is an emergent symmetry that is present because the Heisenberg interaction contains only terms such as $\hat{\bm{S}}_{\bm{R}_i}\cdot \hat{\bm{S}}_{\bm{R}_j}$, which is invariant under the permutation of indices $i$ and $j$. \textit{The $C_{2y}'$ symmetry referred to in the main text is the $\hat{\cal C}_{2y}'$ symmetry}, and it is present in the magnon-phonon coupling Hamiltonian, in contrast to $\hat{{\cal C}}_{2y}$, which is broken by the magnon-phonon coupling Hamiltonian.
Before going further, let us note that the transformation of the magnon operator $\hat{a}_{\bm{R}_i}$ are fixed by Eq.~\eqref{eq.local_spin_transformation} to be 
\begin{equation}
\hat{{\cal C}}  \hat{a}_{\bm{R}_i}\hat{{\cal C}}^{-1} = \hat{a}_{C\bm{R}_i}.\label{eq.hp_transformation}
\end{equation}

Let us next find symmetry operators that acts exclusively on HP operators, corresponding to the symmetries defined by Eq.~\eqref{eq.local_axes_transformation} and \eqref{eq.local_spin_transformation}. 
In other words, we would like to know whether we can absorb the transformation of the local axes into the transformation of HP operators, so that we can treat the local axes as numbers instead of operators.
This step is necessary because when we write the magnon Hamiltonian, as well as the magnon-phonon coupling Hamiltonian, quantities such as $\hat{\bm{x}}_i \cdot \hat{\bm{x}}_j$ will be evaluated to a number which do not transform under the symmetry operator that acts only on the HP operators. 
We will label such a symmetry operator acting only on the HP operators with a superscript`$HP$', $\hat{\cal C}^{HP}$.
To find the action of  $\hat{\cal C}^{HP}$, it is useful to explicitly write down $\hat{\bm{S}}_1 \cdot \hat{\bm{S}}_2$ without numerically evaluating the local axes. Using
\begin{widetext}
\begin{equation}
\hat{\bm{S}}_i=\hat{\bm{x}}_i \frac{\sqrt{2S}}{2} (\hat{a}_i+\hat{a}_i^\dagger)+\hat{\bm{y}}_i \frac{\sqrt{2S}}{2i} (\hat{a}_i-\hat{a}_i^\dagger)+\hat{\bm{z}}_i (S-\hat{a}_i^\dagger \hat{a}_i),
\end{equation}
we find that up to the quadratic order in HP operators, 
\begin{align}
\hat{\bm{S}}_1 \cdot \hat{\bm{S}}_2=&\frac{S}{2}\bigg[
\hat{\bm{x}}_1\cdot \hat{\bm{x}}_2(\hat{a}_1+{a}_1^\dagger)(\hat{a}_2+\hat{a}_2^\dagger)-
2i\hat{\bm{x}}_1\cdot \hat{\bm{y}}_2(\hat{a}_1^\dagger \hat{a}_2-\hat{a}_1 \hat{a}_2^\dagger)-
\hat{\bm{y}}_1\cdot \hat{\bm{y}}_2(\hat{a}_1-{a}_1^\dagger)(\hat{a}_2-\hat{a}_2^\dagger)
-2\hat{\bm{z}}_1 \cdot \hat{\bm{z}}_2(\hat{a}_1^\dagger \hat{a}_1+\hat{a}_2^\dagger \hat{a}_2)
\bigg], \label{eq.s1s2_HP}
\end{align}
where we have used $\hat{\bm{x}}_1\cdot \hat{\bm{y}}_2=-\hat{\bm{y}}_1\cdot \hat{\bm{x}}_2$. Similar expressions for $\hat{\bm{S}}_2 \cdot \hat{\bm{S}}_3$ and $\hat{\bm{S}}_3 \cdot \hat{\bm{S}}_1$ can be obtained through cyclic permutation. 

As an example, let us first discuss $\hat{\cal C}_{3z}$ operator. Because we assume that there is no canting, $\hat{\bm{x}}_1\cdot \hat{\bm{y}}_2=0$. 
The action of $\hat{\cal C}_{3z}$ is
\begin{align}
\hat{\cal C}_{3z} \hat{\bm{S}}_1 \cdot \hat{\bm{S}}_2 \hat{\cal C}_{3z}^{-1}=&\frac{S}{2}\bigg[
\hat{\bm{x}}_2\cdot \hat{\bm{x}}_3(\hat{a}_2+{a}_2^\dagger)(\hat{a}_3+\hat{a}_3^\dagger)-
\hat{\bm{y}}_2\cdot \hat{\bm{y}}_3(\hat{a}_2-{a}_2^\dagger)(\hat{a}_3-\hat{a}_3^\dagger)
-2\hat{\bm{z}}_2 \cdot \hat{\bm{z}}_3(\hat{a}_2^\dagger \hat{a}_2+\hat{a}_3^\dagger \hat{a}_3)
\bigg] \nonumber \\
\overset{(*)}{=}&\frac{S}{2}\bigg[
\hat{\bm{x}}_1\cdot \hat{\bm{x}}_2(\hat{a}_2+{a}_2^\dagger)(\hat{a}_3+\hat{a}_3^\dagger)-
\hat{\bm{y}}_1\cdot \hat{\bm{y}}_2(\hat{a}_2-{a}_2^\dagger)(\hat{a}_3-\hat{a}_3^\dagger)
-2\hat{\bm{z}}_1 \cdot \hat{\bm{z}}_2(\hat{a}_2^\dagger \hat{a}_2+\hat{a}_3^\dagger \hat{a}_3)
\bigg].
\end{align}
For the first equality, we have used Eqs.~\eqref{eq.local_axes_transformation} and \eqref{eq.hp_transformation}, and$\overset{(*)}{=}$ means that the expressions are numerically equivalent.
Because similar equality holds for other pairs of spins, there is a HP operator representation of $\hat{\cal C}_{3z}^{HP}$, which acts only on the HP operators, as a permutation of HP operators:
\begin{equation}
\hat{\cal C}_{3z}^{HP} \hat{a}_{\bm{R}_i} \hat{\cal C}_{3z}^{HP-1}=\hat{a}_{C_{3z}\bm{R}_i}.
\end{equation}
Similarly, we may define
\begin{equation}
\hat{\cal T}_{\bm{R}}^{HP} \hat{a}_{\bm{R}_i} \hat{\cal T}_{\bm{R}}^{HP-1}=\hat{a}_{\bm{R}_i+\bm{R}},\quad \hat{\cal C}_{2x}^{HP} \hat{a}_{\bm{R}_i} \hat{\cal C}_{2x}^{HP-1}=\hat{a}_{C_{2x}\bm{R}_i}, \quad \hat{\cal C}_{2y}^{HP} \hat{a}_{\bm{R}_i} \hat{\cal C}_{2y}^{HP-1}=\hat{a}_{C_{2y}\bm{R}_i}, \quad \hat{\cal C}_{2y}^{'HP} \hat{a}_{\bm{R}_i} \hat{\cal C}_{2y}^{'HP-1}=\hat{a}_{C_{2y}\bm{R}_i}. \label{eq.symmetries_hp}
\end{equation}
Let us note that $\hat{\cal C}_{2y}^{HP}$ is not present in magnon-phonon coupling Hamiltonian, and the definition of $\hat{\cal C}_{2y}^{'HP}$ needs to be modified, as explained in Appendix~\ref{ap.mp_symmetry}. 

Let us next find the representation in the $\bm{k}$ space. For any of the symmetry operators $\hat{\cal C}$,
\begin{equation}
\hat{\cal C}^{HP} \hat{a}_{\bm{k}} \hat{\cal C}^{HP-1}=\sum_{\bm{R}_i}e^{i\bm{k}\cdot \bm{R}_i} \hat{a}_{C\bm{R}_i}=\sum_{\bm{R}_i}e^{i\bm{k}\cdot C^{-1}\bm{R}_i} \hat{a}_{\bm{R}_i}.
\end{equation}
Therefore, when $\hat{\cal C}$ is one of the rotation operators, we have $\hat{\cal C}^{HP}\hat{a}_{\bm{k}} \hat{\cal C}^{HP-1}=\hat{a}_{C\bm{k}}$. When $\hat{\cal C}={\cal T}_{\bm{R}}$, $\hat{\cal C}^{HP}\hat{a}_{\bm{k}} \hat{\cal C}^{HP-1}=e^{-i\bm{k}\cdot \bm{R}}\hat{a}_{\bm{k}}$.
It follows that when $\hat{\cal C}$ is one of the rotation operators, its constraint on the magnon Hamiltonian is $H_m(C\bm{k})=H_m(\bm{k})$.
At this point, there is no difference between ${\cal \hat{C}}_{2y}$ and ${\cal \hat{C}}_{2y}'$. 
This is not so when there is canting, which we explain below, and also when we consider the coupling between magnon and phonon in Appendix~\ref{ap.mp_symmetry}.
Finally, let us note that anisotropy term does not affect the symmetry of the Hamiltonian.
\end{widetext}

\begin{figure}[t]
\centering
\includegraphics[width=8cm]{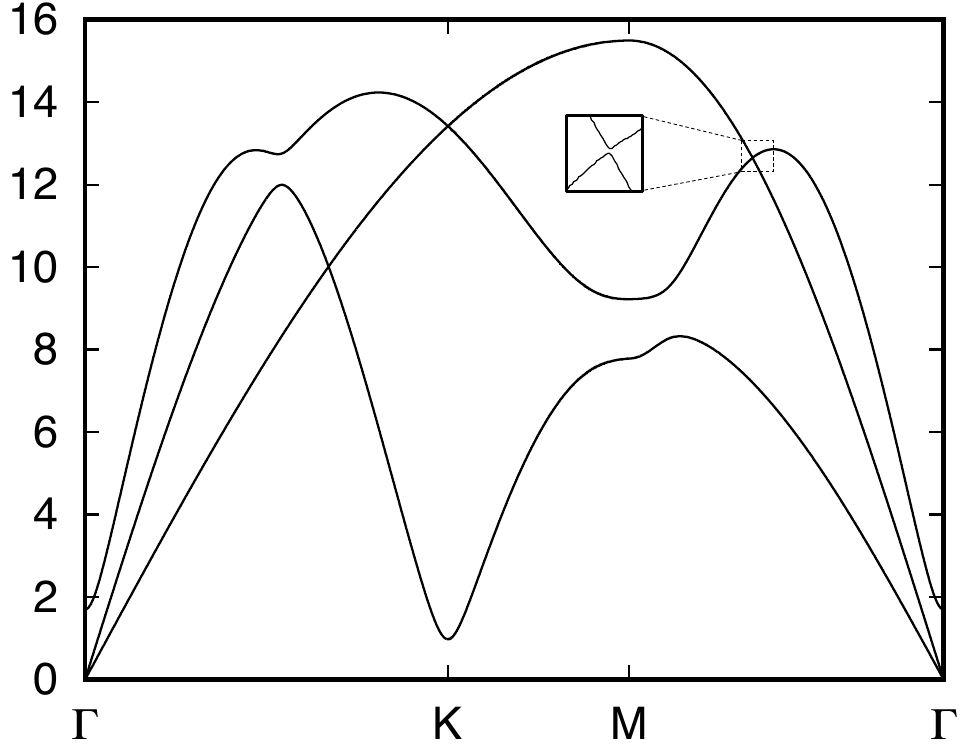}
\caption{Magnon-polaron spectrum with $h=0$ and $H\neq 0$ with the other parameters are the same as in Fig.~\ref{fig.mp_hsl} (d). The band gap along the $M\Gamma$ line opens because $\hat{C}_{2y}'$ is broken, in contrast to the band crossing along the $\Gamma K$ line protected by $\hat{C}_{2x}$.}
\label{fig.mp_hsl_symmetry}
\end{figure}

Let us next comment on what happens when the magnetic order cants. 
As before, ${\cal \hat{T}}_{\bm{R}}$, ${\cal \hat{C}}_{3z}$ and ${\cal \hat{C}}_{2x}$ remain a good symmetry, and their corresponding symmetry acting exclusively on the HP operators do not change.
However, ${\cal \hat{C}}_{2y}$ does not remain a symmetry because the ground state configuration is not invariant under this symmetry. On the other hand, ${\cal \hat{C}}_{2y}'$ is still a symmetry of the Hamiltonian, which can be seen from Eq.~\eqref{eq.s1s2_HP}. 
However, we cannot define a unitary ${\cal \hat{C}}_{2y}'^{HP}$ symmetry.
To see this, notice that the only additional term compared to the case without canting is $iS\hat{\bm{x}}_1\cdot \hat{\bm{y}}_2(\hat{a}_2^\dagger \hat{a}_1-\hat{a}_2\hat{a}_1^\dagger)$. Its transformation is
\begin{align}
{\cal \hat{C}}_{2y}'iS\hat{\bm{x}}_1\cdot \hat{\bm{y}}_2(\hat{a}_2^\dagger \hat{a}_1-\hat{a}_2\hat{a}_1^\dagger) {\cal \hat{C}}_{2y}^{'-1}
&= iS\hat{\bm{x}}_2\cdot \hat{\bm{y}}_1(\hat{a}_1^\dagger \hat{a}_2-\hat{a}_1\hat{a}_2^\dagger) \nonumber \\
&\overset{(*)}{=}iS\hat{\bm{x}}_1\cdot \hat{\bm{y}}_2(\hat{a}_2^\dagger \hat{a}_1-\hat{a}_2\hat{a}_1^\dagger).
\end{align} 
However, we cannot define unitary transformation $\hat{\cal C}_{2y}^{'HP}$ that acts only on the HP operators. 
This is because if we define 
\begin{equation}
\hat{C}^{'HP}_{2y} \hat{a}_{1(2)} \hat{C}^{'HP-1}_{2y}=e^{i\phi}\hat{a}_{2(1)}, \label{eq.tentative_action}
\end{equation}
we have
\begin{align}
{\cal \hat{C}}_{2y}^{'HP}iS\hat{\bm{x}}_1\cdot \hat{\bm{y}}_2&(\hat{a}_2^\dagger \hat{a}_1-\hat{a}_2\hat{a}_1^\dagger) {\cal \hat{C}}_{2y}^{'HP-1} \nonumber \\
&=-iS\hat{\bm{x}}_1\cdot \hat{\bm{y}}_2(\hat{a}_2^\dagger \hat{a}_1-\hat{a}_2\hat{a}_1^\dagger) ,
\end{align}
so that we cannot define a unitary ${\cal \hat{C}}_{2y}^{'HP}$. 
Because the symmetry that protects the gap closing point between the magnon and the phonon bands along the $M\Gamma$ line is unitary $\hat{C}_{2y}^{'HP}$ symmetry, which is also present in the magnon-phonon Hamiltonian when there is no canting  [see Appendix~\ref{ap.mp_symmetry}], the gap along this line will open once the magnetic field is introduced into the magnon Hamiltonian (but not into the phonon Hamiltonian).
 In contrast, the $\hat{C}_{2x}$ symmetry is retained in the presence of canting, and the gap along the $\Gamma K$ line does not open unless magnetic field is introduced into the phonon Hamiltonian as well. This can be seen from the band structure shown in Fig.~\ref{fig.mp_hsl_symmetry}
 
\subsection{Phonon Symmetry} \label{ap.phonon_symmetry}

Let us briefly review the representation of phonon symmetry. Let ${\cal \hat{C}}$ be a spatial symmetry operator and let $C$ be its (vector) representation. Let us use the convention that the displacement vector $\bm{u}$ is a column vector. The potential energy of displacements by $\bm{u}_\alpha (\bm{R}_{i})$ should be equivalent to the potential energy of displacements by $C \bm{u}_{\alpha'} (\bm{R}_{i}') $ when $\bm{u}_\alpha (\bm{R}_{i})=\bm{u}_{\alpha'} (\bm{R}_{i}')$, where the atom at lattice position $\bm{R}_i$ and sublattice $\alpha$ is sent to, under the action of $\hat{C}$, the atom at lattice position $\bm{R}_i'$ and sublattice $\alpha'$. That is,
\begin{align*}
&\bm{u}^T_\alpha (\bm{R}_{i}) K_{\alpha \beta}(\bm{R}_{i}-\bm{R}_{j}) \bm{u}_{\beta}(\bm{R}_{j}) \\
&=[C \bm{u}_{\alpha'} (\bm{R}_{i}')]^TK_{\alpha' \beta'}(\bm{R}_{i}'-\bm{R}_{j}') [C\bm{u}_{\beta'}(\bm{R}_{j}')] \\
&=\bm{u}^T_{\alpha'} (\bm{R}_{i}')[C^T K_{\alpha' \beta'}(\bm{R}_{i}'-\bm{R}_{j}') C] \bm{u}_{\beta'}(\bm{R}_{j}').
\end{align*}
 Thus, we have
\begin{equation}
K_{\alpha' \beta'}(\bm{R}_{i}'-\bm{R}_{j}')=C K_{\alpha \beta}(\bm{R}_{i}-\bm{R}_{j}) C^T.
\end{equation}
The representation of ${\cal \hat{C}}$ consistent with the above is 
\begin{equation}
{\cal \hat{C}} \hat{\bm{u}}_{\alpha}(\bm{R}_i){\cal \hat{C}}^{-1}=C^T \hat{\bm{u}}_{\alpha'}(\bm{R}_{i}'). \label{eq.displ_symmetry}
\end{equation}
 
In the case of triangular lattice without effective magnetic field, the Hamiltonian has   ${\cal \hat{C}}_{3z}$, ${\cal \hat{C}}_{2x}$, and ${\cal \hat{C}}_{2y}'$ symmetries (for phonon, there is no difference between ${\cal \hat{C}}_{2y}$ and ${\cal \hat{C}}_{2y}'$). When acting on $\hat{\bm{u}}(\bm{k})$, their representations in $\bm{k}$ space is
\begin{equation}
C_{3z}=\begin{pmatrix}
-\frac{1}{2} & \frac{\sqrt{3}}{2} \\
-\frac{\sqrt{3}}{2} & -\frac{1}{2}
\end{pmatrix},~
C_{2x}=\begin{pmatrix}
1 & 0 \\
0 & -1
\end{pmatrix},~ C_{2y}'=\begin{pmatrix}
-1 & 0 \\
0 & 1
\end{pmatrix}.
\end{equation}
On the other hand, along the high symmetry lines $\Gamma K$ and $M \Gamma$, the dynamical matrix is diagonal. Along $\Gamma K$, $D_{xx}(\bm{k})-D_{yy}(\bm{k})=\frac{2\gamma}{M}\left[ \cos \frac{k_x}{2}-\cos k_x\right]>0$ so that the band with higher (lower) energy has $C_{2x}$ eigenvalue of $1(-1)$. Similarly, along the $M\Gamma$ line, 
\begin{equation}
D_{xx}-D_{yy}=\frac{2\gamma}{M}\left[ \cos \frac{\sqrt{3}k_y}{2}-1\right]<0,
\end{equation}
so that the band with higher (lower) energy has $C_{2y}'$ eigenvalue of $1(-1)$.

\subsection{Magnon-Phonon Symmetry} \label{ap.mp_symmetry}
\begin{widetext}
The magnon-phonon coupling Hamiltonian does not break any of the symmetries we have mentioned above except for  $\hat{\cal C}_{2y}$, and the technique we have used to find the symmetry representations of magnon and phonon can be straightforwardly applied even in the presence of magnon-phonon coupling, except for $\hat{\cal C}_{2y}'$. 
Therefore, we will focus on $\hat{\cal C}_{2y}$  and $\hat{\cal C}_{2y}'$ symmetries.
To understand how $\hat{\cal C}_{2y}$  and $\hat{\cal C}_{2y}'$ acts on the coupling Hamiltonian, let us restore explicitly write the local axes in the expression for $\hat{\bm{S}}_1\cdot \hat{\bm{S}}_2$ that is linear in HP operators:
\begin{align}
\hat{\bm{S}}_1 \cdot \hat{\bm{S}}_2=&\frac{S\sqrt{2S}}{2}\bigg[ (\hat{a}_1+\hat{a}_1^\dagger)\hat{\bm{x}}_1\cdot \hat{\bm{z}}_2-i(\hat{a}_1-\hat{a}_1^\dagger)\hat{\bm{y}}_1\cdot \hat{\bm{z}}_2 -i(\hat{a}_2-\hat{a}_2^\dagger)\hat{\bm{z}}_1\cdot \hat{\bm{y}}_2+(\hat{a}_2+\hat{a}_2^\dagger)\hat{\bm{z}}_1\cdot \hat{\bm{x}}_2\bigg]. \label{eq.s1s2_to_linear}
\end{align}
Using Eq.~\eqref{eq.local_axes_transformation} and \eqref{eq.hp_transformation}, 
\begin{align}
\hat{\cal C}_{2y} \hat{\bm{S}}_1 \cdot \hat{\bm{S}}_2 \hat{\cal C}_{2y}^{-1} &= \frac{S\sqrt{2S}}{2}\bigg[ (\hat{a}_2+\hat{a}_2^\dagger)(-\hat{\bm{x}}_2)\cdot \hat{\bm{z}}_1-i(\hat{a}_2-\hat{a}_2^\dagger)(-\hat{\bm{y}}_2)\cdot \hat{\bm{z}}_1 -i(\hat{a}_1-\hat{a}_1^\dagger)\hat{\bm{z}}_2\cdot (-\hat{\bm{y}}_1)+(\hat{a}_1+\hat{a}_1^\dagger)\hat{\bm{z}}_2\cdot (-\hat{\bm{x}}_1)\bigg] \nonumber \\
& = -\hat{\bm{S}}_1\cdot \hat{\bm{S}}_2 \neq \hat{\bm{S}}_1\cdot \hat{\bm{S}}_2 \label{eq.s1s2__linear_c2y} .
\end{align}
It follows from this that $\hat{\cal C}_{2y} $ is not a symmetry of the magnon-phonon coupling Hamiltonian even in the absence of canting because
\begin{align}
{\cal \hat{C}}_{2y} \hat{\bm{S}}_1\cdot \hat{\bm{S}}_2 \bm{R}_{12}\cdot (\hat{\bm{u}}_1-\hat{\bm{u}}_2){\cal \hat{C}}_{2y}^{-1}= (-\hat{\bm{S}}_1\cdot \hat{\bm{S}}_2)\bm{R}_{12}\cdot (\hat{\bm{u}}_1-\hat{\bm{u}}_2),
\end{align}
where we used Eqs.~\eqref{eq.displ_symmetry} and \eqref{eq.s1s2__linear_c2y}.
On the other hand, 
\begin{align}
\hat{\cal C}_{2y}' \hat{\bm{S}}_1 \cdot \hat{\bm{S}}_2 \hat{\cal C}_{2y}^{'-1} &= \frac{S\sqrt{2S}}{2}\bigg[ (\hat{a}_2+\hat{a}_2^\dagger)\hat{\bm{x}}_2\cdot \hat{\bm{z}}_1-i(\hat{a}_2-\hat{a}_2^\dagger)\hat{\bm{y}}_2\cdot \hat{\bm{z}}_1 -i(\hat{a}_1-\hat{a}_1^\dagger)\hat{\bm{z}}_2\cdot \hat{\bm{y}}_1+(\hat{a}_1+\hat{a}_1^\dagger)\hat{\bm{z}}_2\cdot \hat{\bm{x}}_1)\bigg] \nonumber \\
& = \hat{\bm{S}}_1\cdot \hat{\bm{S}}_2 ,
\end{align}
so that $\hat{\cal C}_{2y}' $ is a symmetry of the coupling Hamiltonian. 
However, we must still be careful because the ${\cal \hat{C} }_{2y}^{HP}$ symmetry operator that acts only on the HP operators as ${\cal \hat{C} }_{2y}^{HP} a_{\bm{R}_i} {\cal \hat{C} }_{2y}^{'HP-1}= a_{C_{2y}\bm{R}_i} $ in Eq.~\eqref{eq.symmetries_hp} does not carry over, as we now explain. 
Because $\hat{\cal C}_{2y}' $ is not unitary-representable in the magnon Hamiltonian when there is canting, we will only treat the case without canting. Using $\hat{\bm{x}}_1\cdot \hat{\bm{z}}_2=-\hat{\bm{x}}_2\cdot \hat{\bm{z}}_1$ and $\hat{\bm{z}}_1\cdot \hat{\bm{y}}_2=\hat{\bm{z}}_2\cdot \hat{\bm{y}}_1=0$,
\begin{align}
\hat{\cal C}_{2y}' \hat{\bm{S}}_1 \cdot \hat{\bm{S}}_2 \hat{\cal C}_{2y}^{'-1} &\overset{(*)}{=}- \frac{S\sqrt{2S}}{2}\bigg[ (\hat{a}_2+\hat{a}_2^\dagger)\hat{\bm{x}}_1\cdot \hat{\bm{z}}_2+(\hat{a}_1+\hat{a}_1^\dagger)\hat{\bm{z}}_1\cdot \hat{\bm{x}}_2)\bigg] \nonumber \\
&=\hat{\cal C}_{2y}^{'HP}   \frac{S\sqrt{2S}}{2}\bigg[ (\hat{a}_1+\hat{a}_2^\dagger)\hat{\bm{x}}_1\cdot \hat{\bm{z}}_1+(\hat{a}_2+\hat{a}_2^\dagger)\hat{\bm{z}}_1\cdot \hat{\bm{x}}_2)\bigg]  \hat{\cal C}_{2y}^{'HP-1} =\hat{\cal C}_{2y}^{'HP} \hat{\bm{S}}_1 \cdot \hat{\bm{S}}_2 \hat{\cal C}_{2y}^{'HP-1} ,
\end{align}
if we (re)-define
\begin{equation}
\hat{\cal C}_{2y}^{'HP} \hat{a}_{\bm{R}_i} \hat{\cal C}_{2y}^{'HP-1} =-\hat{a}_{C_{2y}\bm{R}_i}.
\end{equation}
Therefore, we see that the eigenvalue of $\hat{\cal C}_{2y}^{'HP}$ is $-1$ along the $M\Gamma$ line for the magnon band.
On the other hand, $\hat{\cal C}_{2x}^{HP}$ defined as in Eq.~\eqref{eq.symmetries_hp} remains valid and its eigenvalue along the $\Gamma K$ line is $1$. 
\end{widetext}

\section{Diagonalization of BdG Hamiltonian} \label{ap.BdG_review}

In this section, we summarize Colpa's method\cite{colpa1978diagonalization}  of  finding a matrix $T$ that diagonalizes a  $2N\times 2N$ positive definite bosonic BdG Hamiltonian $H_{\textrm{BdG}}$.  Then we show that when the eigenvalues are non-degenerate, the matrix $T$ satisfy a uniqueness condition.
To diagonalize $H_{BdG}$, we must find $T$ that satisfies 
\begin{equation}
T^\dagger H_{\textrm{BdG}} T=\textrm{diag}(\tilde{E}_{n}),\quad T^\dagger \sigma_z T. 
\end{equation}
First, make the decomposition $H_{BdG}=K^\dagger K$, which can be numerically implemented by the Cholesky decomposition. 
Second, define $U$ to be the matrix that unitarily diagonalizes $K \sigma_z K^\dagger$: $U^\dagger [K \sigma_z K^\dagger]U=E$. Here, $E$ is a diagonal matrix with $N$ positive and $N$ negative entries. Then, $\tilde{E}=\sigma_z E$ and $T=K^{-1} U \sqrt{\tilde{E}}$.  
Let us also note that when $H_{BdG}$ is real, $K$ can be taken to be real, so that $T$ is real, i.e. the Hamiltonian can be diagonalized by a real matrix.

Let us define $|T,n \rangle$ to be the $n$th column of $T$, and refer to it as an eigenvector of $H_{BdG}$. We will now show that it is unique up to an overall phase factor when there is no degeneracy in energy spectrum. 
It follows that there is no problem in assuming that the eigenvectors obtained from directly diagonalizing $H_s$ defined in Eq.~\eqref{eq.simplified_hamiltonian} is given by Eq.~\eqref{eq.simplified_eigenvector} when we compute the Berry curvature.

Let $T$ and $\tilde{T}$ be two sets of matrices that diagonalizes $H_{BdG}$.
Because $\textrm{det}(T)\neq 0$, we can write any vector as a linear combination of  $|T,m \rangle$. Therefore,
\begin{align}
&|\tilde{T},n \rangle=\sum_{m} |T,m\rangle C_{nm}\nonumber \\
\Leftrightarrow &  H_{BdG} |\tilde{T},n\rangle=\sum_{m} H_{BdG}|T,m\rangle C_{nm} \nonumber \\
\Leftrightarrow& E_n \sigma_z  |\tilde{T},n\rangle=\sum_m E_m \sigma_z |T,m\rangle C_{nm} \nonumber \\
\Leftrightarrow& E_n \langle T,n' | \sigma_z | \tilde{T},n\rangle=\sum_{m} E_m \langle T,n' | \sigma_z | T,m\rangle C_{nm} \nonumber \\
\Leftrightarrow &E_n \langle T,n' | \sigma_z | \tilde{T},n\rangle= E_{n'} (\sigma_z)_{n'n'}C_{nn'}.
\end{align}
To obtain the third line, we use $H_{BdG}|T,n\rangle=E_n|T,n\rangle$  and $H_{BdG}|\tilde{T},n\rangle=E_n|\tilde{T},n\rangle$.
On the other hand, 
\begin{align}
\langle T,n' | H_{BdG} | \tilde{T}, n\rangle &= E_n \langle T,n' | \sigma_z | \tilde{T},n\rangle \nonumber \\
&= E_{n'} \langle T,n' | \sigma_{z} | \tilde{T},n\rangle,
\end{align}
where we have used $\langle T,n'|H_{BdG}=E_{n'}\langle T,n'| \sigma_z$. This implies that $\langle T,n' | \sigma_{z} | \tilde{T},n\rangle$ is nonzero iff $n=n'$.
From the above two equations, we see that $C_{nn'}$ is nonzero iff $n=n'$. Therefore, $|\tilde{T},n\rangle=C_{nn}|T,n\rangle$ where $C_{nn}$ is a phase factor. This concludes the proof.

\section{Magnetostriction}

Because magnetoelastic coupling can lead to magnetostriction, we should check whether magnetostriction will occur for the model we have considered in the main text. 
We show below that there will be no magnetostriction at the mean field level. 
To see this, let us first focus on the magnetoelastic coupling between sites 1 and 2 in Fig.~\ref{fig.triangular_lattice}. 
The term that can potentially cause magnetostriction is 
\begin{equation}
J\langle \bm{S}_{1} \rangle \cdot \langle \bm{S}_{2} \rangle (u_{2x}-u_{1x}). \label{eq.h_magnetostriction1}
\end{equation}
Similarly, if we consider the magnetoelastic coupling between sites 11 and 1, we obtain \begin{equation}
J\langle \bm{S}_{11} \rangle \cdot \langle \bm{S}_{1} \rangle (u_{1x}-u_{11x}). \label{eq.h_magnetostriction2}
\end{equation}
We thus see that terms proportional to $u_{1x}$ cancel in Eqs.~\eqref{eq.h_magnetostriction1} and \eqref{eq.h_magnetostriction2}. Similar cancellation occurs in magnetoelastic coupling between other sites, so the model we used in the main text does not cause magnetostriction at the mean field level.

However, if we consider magnetoelastic coupling arising from spin-orbit coupling, there can be magnetostriction. Magnetoelastic coupling arising from spin-orbit coupling  is quadratic in magnetization and linear in strain tensor \cite{gurevich1996magnetization}, and as an example, we can write down the following term for the magnetoelastic coupling between sites 1 and 2 in Fig.~\ref{fig.triangular_lattice},
\begin{equation}
\alpha s_{1x}s_{1y}(u_{2y}-u_{1y}).
\end{equation}
Here, $s_{ix}=\bm{S}_{i}\cdot \hat{\bm{x}}$, $s_{iy}=\bm{S}_i\cdot \hat{\bm{y}}$, and $\hat{\bm{x}}$ and $\hat{\bm{y}}$ are unit vectors along \textit{global} x and y axis. From this, we can find the coupling between the others by imposing triangular lattice symmetry. Such a term will cause magnetostriction because terms linear in $u_y$ are not cancelled. This result is reasonable because the magnetic order breaks the translation symmetry, and it should be expected that the lattice will be able to see this through spin-orbit coupling. For simplicity of the model, we will not consider such terms. 

\section{Phonon Conventions} \label{ap.other_formulation}
Let us first relate the phonon normalization given in the main text with that given in Refs.~[\onlinecite{{zhang2011phonon,zhang2010topological}}].
There, the authors define right eigenvectors $|\bm{\chi}^R_{\bm{k},\sigma}\rangle$ and left eigenvectors $|\bm{\chi}^L_{\bm{k},\sigma}\rangle$ of $H_{\textrm{eff}}$. 
Let us define 
\begin{equation}
|\bm{\chi}^R_{\bm{k},\sigma}\rangle=\sqrt{2|\omega_{\bm{k},\sigma}|}|\bm{\chi}_{\bm{k},\sigma}\rangle,
\end{equation}
where $|\bm{\chi}_{\bm{k},\sigma}\rangle$ is a right eigenvector [see Eq.~\eqref{eq.phonon_eigenvector}] satisfying the normalization condition in Eq.~\eqref{eq.chi_normalization}. 
Then, it is easy to check that  
\begin{equation}
|\bm{\chi}^L_{\bm{k},\sigma}\rangle=\frac{\rho_y}{2\omega_{\bm{k},\sigma}}|\bm{\chi}^R_{\bm{k},\sigma}\rangle
\end{equation}
 is a left eigenvector of $H_{\textrm{eff}}$. 
 Since we have normalized $|\bm{\chi}_{\bm{k},\sigma}\rangle$ using $\rho_y$ as in Eq.~\eqref{eq.chi_normalization}, we see that the right and left eigenvectors we have defined satisfy the normalization condition given in Refs.~[\onlinecite{{zhang2011phonon,zhang2010topological}}],
\begin{equation}
\langle \bm{\chi}^{L}_{\bm{k},\sigma}|\bm{\chi}^{R}_{\bm{k},\sigma'}\rangle=\delta_{\sigma,\sigma'}. \label{eq.lr_normalization}
\end{equation}
If we define
\begin{equation}
\bm{\bm{\chi}}^R_\sigma(\bm{k})=\begin{pmatrix}
\bm{\mu}^R_{\sigma} (\bm{k})\\
\bm{\epsilon}^R_{\sigma} (\bm{k})
\end{pmatrix}  \label{eq.polarization},
\end{equation}
Eq.~\eqref{eq.lr_normalization} becomes
\begin{equation}
\langle \bm{\chi}^{L}_{\bm{k},\sigma}|\bm{\chi}^{R}_{\bm{k},\sigma'}\rangle=\bm{\epsilon}^{R\dagger}_{\bm{j},\sigma} \bm{\epsilon}^R_{\bm{j},\sigma'}+\frac{i}{\omega_{\bm{k},\sigma}}\bm{\epsilon}^{R\dagger}_{\bm{j},\sigma} A\bm{\epsilon}^R_{\bm{j},\sigma'}=\delta_{\sigma,\sigma'} \label{eq.rightpol_normalization},
\end{equation}
where we have used Eq.~\eqref{eq.polarization_relation}.

It is not difficult to show why this normalization is possible. First, let us note that when there are no band degeneracies, left and right eigenvectors are orthogonal, so that $\langle \bm{\chi}^{L}_{\bm{k},\sigma}|\bm{\chi}^{R}_{\bm{k},\sigma'}\rangle\propto \delta_{\sigma,\sigma'}$ (we have not normalized the left and right eigenvectors at this point).
We then notice from Eq.~\eqref{eq.phonon_eigenvector} that
\begin{equation}
(-\omega_{\bm{k},\sigma}^2-2i\omega_{\bm{k},\sigma}A+A^2+D)\bm{\epsilon}^R_{\sigma} (\bm{k})=0 \label{eq.rightpol_eigenvector}.
\end{equation}
Multiplying  Eq.~\eqref{eq.rightpol_eigenvector} by $\bm{\epsilon}^{R\dagger}_{\bm{j},\sigma} $ to the left, we obtain
\begin{equation}
\bm{\epsilon}^{R\dagger}_{\bm{j},\sigma} \bm{\epsilon}^R_{\bm{j},\sigma'}+\frac{i}{\omega_{\bm{k},\sigma}}\bm{\epsilon}^{R\dagger}_{\bm{j},\sigma} A\bm{\epsilon}^R_{\bm{j},\sigma'}=\bm{\epsilon}^{R\dagger}_{\bm{j},\sigma}\left(\frac{1}{2}+\frac{A^2+D(\bm{k})}{2\omega_{\bm{k},\sigma}^2} \right) \bm{\epsilon}^R_{\bm{j},\sigma'}.
\end{equation}
Using Eq.~\eqref{eq.dynamical_matrix}, we see that the matrix enclosed in parenthesis on the right hand side of the above equation is positive definite. Therefore, $\langle \bm{\chi}^{L}_{\bm{k},\sigma}|\bm{\chi}^{R}_{\bm{k},\sigma'}\rangle=\langle \bm{\chi}^{L}_{\bm{k},\sigma}|\bm{\chi}^{R}_{\bm{k},\sigma}\rangle\delta_{\sigma,\sigma'}$
where $\langle \bm{\chi}^{L}_{\bm{k},\sigma}|\bm{\chi}^{R}_{\bm{k},\sigma}\rangle > 0$.
Redefining $|\bm{\chi}^R_{\bm{k},\sigma}\rangle \rightarrow \frac{1}{\sqrt{\langle \bm{\chi}^L_{\bm{k},\sigma}|\bm{\chi}^R_{\bm{k},\sigma}\rangle} }|\bm{\chi}^R_{\bm{k},\sigma}\rangle $, we obtain the normalization conditions in 
Eqs.~\eqref{eq.rightpol_normalization}, \eqref{eq.lr_normalization}, and \eqref{eq.chi_normalization}.

Next, let us show that the Berry curvature that we have defined is equivalent to that defined in Refs.~[\onlinecite{{zhang2011phonon,zhang2010topological}}].  The Berry curvature $\bm{B}'_{\bm{k},\sigma}$ defined in Refs.~[\onlinecite{{zhang2011phonon,zhang2010topological}}] is given by
\begin{align}
\bm{B}'_{\bm{k},\sigma}&\equiv\bm{\nabla} \times \langle \bm{\chi}^L_{\bm{k},\sigma}|\bm{\nabla}|\bm{\chi}^R_{\bm{k},\sigma}\rangle \nonumber \\
&= \bm{\nabla} \times \bm{A}_{\bm{k},\sigma}+\bm{\nabla}\times [\frac{2\omega_{\bm{k},\sigma}}{\sqrt{2|\omega_{\bm{k},\sigma}|}}(\sigma_{z})_{\sigma\sigma}\bm{\nabla} \sqrt{2\omega_{\bm{k},\sigma}}] \nonumber \\
&=\bm{B}_{\bm{k},\sigma}.
\end{align}
where 
\begin{align}
\bm{A}_{\bm{k},\sigma}&=i(\rho_{z})_{\sigma \sigma}\langle \bm{\xi}_{\bm{k},\sigma}|\rho_{z}\bm{\nabla}|\bm{\xi}_{\bm{k},\sigma} \rangle \nonumber \\
&=i(\rho_{z})_{\sigma \sigma}\langle \bm{\chi}_{\bm{k},\sigma}|  \rho_{y} \bm{\nabla}|\bm{\chi}_{\bm{k},\sigma} \rangle  \nonumber \\
\bm{B}_{\bm{k},\sigma}&=\bm{\nabla}\times \bm{A}_{\bm{k},\sigma}. \label{eq.phonon_berry}
\end{align}
are the phonon Berry connection and Berry curvature we have defined in the main text.
As discussed in the main text, the BdG nature of phonon implies that for the purpose of calculating the thermal Hall conductivity, it is sufficient to limit ourselves only to the sector for which $n>0$.

Finally, let us note that the convention we use in the main text is convenient for defining Berry curvature, but we must be careful because the polarization vector for position, which appears in magnon-phonon coupling Hamiltonian, is unit dependent in our convention. Using Eq.~\eqref{eq.polarization_relation}, we have
\begin{equation}
\bm{\chi}_{\bm{k},\sigma}^\dagger \sigma_y \bm{\chi}_{\bm{k},\sigma}=2\omega_{\bm{k},\sigma} \bm{\epsilon}_{\bm{k},\sigma}^\dagger \bm{\epsilon}_{\bm{k},\sigma}+2i\bm{\epsilon}_{\bm{k},\sigma}^\dagger A \bm{\epsilon}_{\bm{k},\sigma}.
\end{equation}
Thus, the unit of $\bm{\epsilon}_{\bm{k},\sigma}$ is $[s]^{1/2}$ and the unit of $\mu$ is $[s]^{-1/2}$, in contrast to conventional normalization of the polarization vectors such as that adopted in Refs.~[\onlinecite{{zhang2011phonon,zhang2010topological}}] where $\bm{\epsilon}_{\bm{k},\sigma}$ is unitless and $\mu$ has unit of $[s]^{-1/2}$. Since the unit of $K_{mp}\sqrt{\frac{\hbar}{M}}$  is $[J/s^{1/2}]$,  the unit of $K_{mp}\sqrt{\frac{\hbar}{M}}\bm{\epsilon}_{\bm{k},\sigma}$ is $[J]$. Now, if we put $\hbar=1$ and measure energy in units of meV, we obtain a new unit of time $s'$. If we have a quantity whose unit is $[s]$,  we have the rule $1[s]=f~[s']$. Thus, $ \left(K_{mp}\sqrt{\frac{\hbar}{M}}\frac{1}{1\textrm{meV}}\right) [s']^{-1/2}= \left(K_{mp}\sqrt{\frac{f \hbar}{M}}\frac{1}{1\textrm{meV}}\right) [s]^{-1/2}$, which is the origin of the factor $f$ in Sec.~\ref{ssec.mpctriangular}.

\section{Berry Curvature}  \label{ap.berry}
In this section, we review how the Berry curvature can be computed and explain why the Berry curvature computed from $\tilde{H}_{me}$ does not behave well numerically. Then, we discuss the reality condition on magnon-polaron Berry curvature.
\subsection{Computation of Berry Curvature}
\label{ap.computation}

In the usual system where the particle number is conserved,
the Berry curvature can be calculated by dividing the Brillouin zone into plaquettes and by calculating the flux of the Berry curvature, which is given by \cite{fukui2005chern}:
\begin{align}
\textrm{Arg}[&\langle n \bm{k}  | n \bm{k} +\bm{\delta}_1 \rangle \langle n \bm{k} +\bm{\delta}_1 | n \bm{k} +\bm{\delta}_1 +\bm{\delta}_2\rangle \times \nonumber \\
&\langle n \bm{k}+\bm{\delta}_1+\bm{\delta}_2  | n \bm{k} +\bm{\delta}_2 \rangle\langle n \bm{k} +\bm{\delta}_2 | n \bm{k} \rangle ]. \label{eq.berry_plaquette_nonbdg}
\end{align}
For BdG Hamiltonian, we need to make a slight modification because the projection operator to a set of sub-bands ${\cal S}$ is given by $\sum_{n\in {\cal S}}| n\bm{k}\rangle (\sigma_{z})_{nn}\langle n \bm{k}|\sigma_z$. 
For the purpose of calculating Berry curvature of a single band, it suffices to replace $\langle n \bm{k}|n\bm{k}+\delta \bm{k}\rangle$  by  $\langle n \bm{k}|\sigma_{z}|n\bm{k}+\delta \bm{k}\rangle$ in  Eq.~\eqref{eq.berry_plaquette_nonbdg}, which gives Berry curvature that is equivalent to Eq.~\eqref{eq.berry_curvature} in the main text. 
Namely, the flux of Berry curvature through a plaquette formed by $\bm{k}, \bm{k}+\bm{\delta}_1, \bm{k}+\bm{\delta}_1+\bm{\delta}_{2}, \bm{k}+\bm{\delta}_2$ is given by \cite{shindou2013topological} (modulo $2\pi$)
\begin{align}
\textrm{Arg}[&\langle n \bm{k} | \sigma_z | n \bm{k} +\bm{\delta}_1 \rangle \langle n \bm{k} +\bm{\delta}_1| \sigma_z | n \bm{k} +\bm{\delta}_1 +\bm{\delta}_2\rangle \times \nonumber \\
&\langle n \bm{k}+\bm{\delta}_1+\bm{\delta}_2 | \sigma_z | n \bm{k} +\bm{\delta}_2 \rangle\langle n \bm{k} +\bm{\delta}_2| \sigma_z | n \bm{k} \rangle ] .\label{eq.berry_plaquette}
\end{align}
 This is the method we have used to compute the Berry curvature in the main text.

\begin{figure}[t]
\centering
\includegraphics[width=8cm]{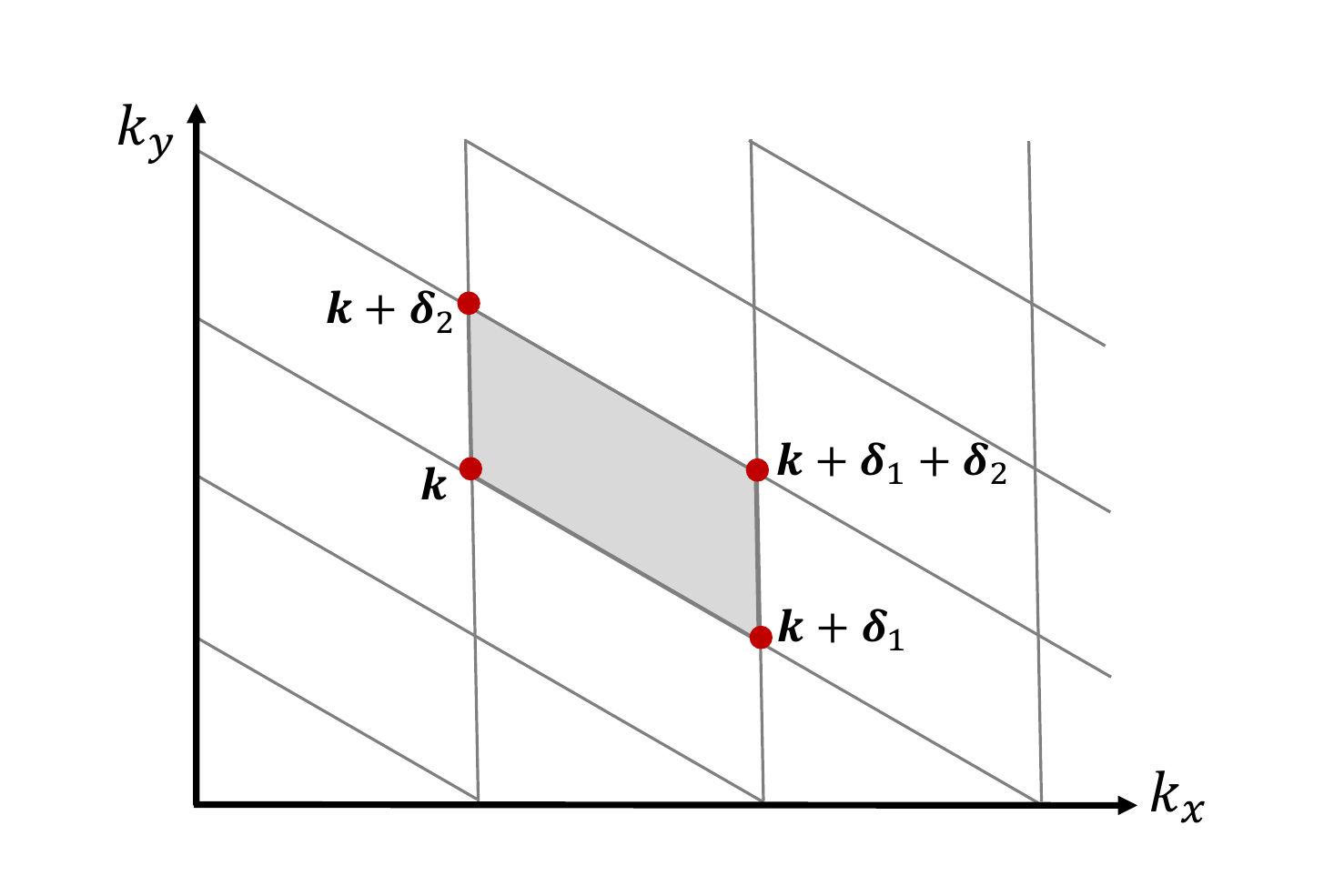}
\caption{Discretized Brillouin zone. The flux of Berry curvature modulo $2\pi$ through the shaded plaquette can be calculated from \eqref{eq.berry_plaquette}.}
\label{fig.plaquette}
\end{figure}

Another way to define a gauge invariant expression for the Berry curvature is to write it in terms of the Hamiltonian. Let us first note the identity:
\begin{equation}
\sum_m \sigma_{z} |m \bm{k}\rangle(\sigma_{z})_{mm} \langle m \bm{k}| \sigma_{z} = \sigma_{z}.
\end{equation}
Thus, the Berry curvature for $n>0$ is 
\begin{equation}
\bm{B}_{n}(\bm{k})=i\sum_{m\neq n} (\bm{\nabla} \langle n\bm{k} |)\sigma_{z} |m \bm{k}\rangle \times (\sigma_{z})_{mm}\langle m\bm{k}| \sigma_{z} \bm{\nabla} | n \bm{k}\rangle.
\end{equation}
Note that the term $m=n$ does not contribute because of the identity  $\langle m \bm{k}| \sigma_z \bm{\nabla}_{\bm{k}} |n\bm{k}\rangle=-(\bm{\nabla}\langle m\bm{k}|)\sigma_{z} |n\bm{k}\rangle$, which follows from taking the gradient of the both sides of $\langle m \bm{k}| \sigma_z |n\bm{k}\rangle=(\sigma_{z})_{mn}$.
The energy eigenstates satisfy
\begin{equation}
H_{\bm{k}}|n\bm{k}\rangle = E_n(\bm{k}) \sigma_{z}|n\bm{k}\rangle,
\end{equation}
where $E_n(\bm{k})$ takes both positive and negative values. Taking the gradient on both sides and multiplying by $\langle m \bm{k} |$, we obtain 
\begin{align}
\langle m\bm{k}| \bm{\nabla} H_{\bm{k}} |n \bm{k}\rangle = (E_n(\bm{k}) -E_m(\bm{k}))\langle m\bm{k}|\sigma_{z} \bm{\nabla} | n \bm{k}\rangle \nonumber \\ 
+\bm{\nabla} E_n (\bm{k})(\sigma_{z})_{mn}.
\end{align}
Thus, 
\begin{equation}
\bm{B}_{n}(\bm{k})=\sum_{m\neq n}\frac{i\langle n\bm{k}| \bm{\nabla} H_{\bm{k}} |m \bm{k}\rangle (\sigma_{z})_{mm} \times \langle m\bm{k}| \bm{\nabla} H_{\bm{k}} |n\bm{k}\rangle}{(E_n(\bm{k})-E_m(\bm{k}))^2}. \label{eq.berry_hamiltonian}
\end{equation}

Let us note that in both of the methods, the Hamiltonian should be smooth. However, when we naively construct the Hamiltonian numerically, $\tilde{H}_{me}$ is not smooth because the phase of the phonon polarization vector $\bm{\epsilon}_{\sigma}(\bm{k})$ is not smoothly determined. The problem this causes in the second method is clear from Eq.~\eqref{eq.berry_hamiltonian}. To clarify what goes wrong in the first method, let us reexamine the toy model in the main text. Let us multiply the polarization vector by some phase factor $\bm{\epsilon}'_{\sigma=1,2} (\bm{k})=e^{-i\zeta_\sigma(\bm{k})} \bm{\epsilon}_\sigma (\bm{k})$. For simplicity, let us assume that $\zeta_\sigma (\bm{k})=\zeta\delta_{\bm{k},\bm{k}_0}-\zeta\delta_{\bm{k},-\bm{k}_0}$. 
Then, the eigenvectors of the bosonic BdG Hamiltonian $\tilde{H}_{me}$ changes to $|n \bm{k}\rangle'=e^{i\zeta(\bm{k})}|n \bm{k}\rangle$ where $e^{-i\zeta(\bm{k})}=\textrm{diag}(1,e^{-i\zeta_1(\bm{k})},e^{-i\zeta_1(\bm{k})},1,e^{i\zeta_1(-\bm{k})},e^{i\zeta_1(-\bm{k})})$. Let us note that only the wave functions at $\bm{k}_0$ and $-\bm{k}_0$  are multiplied by a matrix that is not the identity. When we compute the flux of Berry curvature  through a plaquette containing $\bm{k}_0$, it is clear from Eq.~\eqref{eq.berry_plaquette}  that the flux is not invariant under the transformation $|n \bm{k}\rangle \rightarrow |n \bm{k} \rangle'$.
Let us note that this transformation differs from the usual $U(1)$ transformation of the form $|n \bm{k}\rangle''=|n \bm{k}\rangle e^{i\tilde{\zeta}(\bm{k})}$ where $e^{i\tilde{\zeta}(\bm{k})}$ an overall phase factor multiplying the wavefunction. In this case, it is easily seen that Eq.~\eqref{eq.berry_plaquette} is invariant under the transformation $|n \bm{k}\rangle\rightarrow |n \bm{k}\rangle''$

\subsection{Reality Condition} \label{ap.berry_reality_condition}
Let us first mention that it does not immediately follow that the Berry curvature vanishes from the condition that the matrix $H_{me}(\bm{k})$ is real. For this would imply that the phonon Berry curvature is always be zero after we turn off the magnon-phonon coupling. We will show below that when $h=0$ in phonon Hamiltonian, the reality of $H_{me}(\bm{k})$ implies zero Berry curvature. For notational simplicity, we will omit the dependence on $\bm{k}$ in what follows.

When $h=0$ and there is no magnon-phonon coupling through the phonon momentum $\bm{p}$, $H_{me}$ takes the form
\begin{equation}
H_{me}=\begin{pmatrix}
 H_{m} &\vline& 0 & H_{c} \\
\hline
0 &\vline& \frac{1}{2} I_{sd} & 0 \\
H_{c}^\dagger &\vline& 0 & \frac{1}{2}D
\end{pmatrix}
\end{equation}
[cf. Eq.~\eqref{eq.magnetoelastic_hamiltonian}].  Then, 
\begin{equation}
\tilde{V}^\dagger H_{me} \tilde{V} =
\begin{pmatrix}
\tau_z H_{m} \tau_z  &\vline& \frac{1}{\sqrt{2}}\tau_z H_{c} & -\frac{1}{\sqrt{2}}\tau_z H_{c} \\
\hline
\frac{1}{\sqrt{2}}H_c^\dagger \tau_z &\vline& \frac{1}{4}(1+D) & \frac{1}{4}(1-D) \\
-\frac{1}{\sqrt{2}}H_{c}^\dagger \tau_z &\vline& \frac{1}{4}(1-D) &\frac{1}{4}(1+D)
\end{pmatrix},
\end{equation}
where we have defined $\tilde{V}=[(i\tau_z)\oplus \rho_0] V$, $V$ was defined in the main text and $\rho_0$ is the identity matrix in the phonon sector.  Here, let us note that $\tau_z$ is necessary to keep $H_m$ in BdG form. Then, $\tilde{H}_{s}=P\tilde{V}^\dagger H_{me} \tilde{V} P^\dagger$ is a bosonic BdG Hamiltonian with real components, where the permutation matrix $P$ was defined in Eq.~\eqref{eq.permutation_matrix}. This can be diagonalized with a real matrix  through Colpa's method. Let $|T_n\rangle$ be the $n$th column vector of the real matrix that diagonalizes $\tilde{H}_s$ [see Appendix~\ref{ap.BdG_review}]. Then, 
the (abelian) Berry connection for $n>0$ is 
\begin{align}
i\langle T_n | \sigma_z \bm{\nabla} |T_n \rangle&=i\langle \hat{K} T_n|\sigma_z \bm{\nabla} \hat{K}| T_n \rangle \nonumber \\
&=i [\bm{\nabla} \langle T_n] \sigma_z |T_n \rangle \nonumber \\
&=-i \langle T_n | \bm{\nabla} \sigma_z |T_n\rangle=0,
\end{align} 
where $\hat{K}$ is the complex conjugation operator and we have used $\bm{\nabla}[ \langle T_n |\sigma_z | T_n \rangle]=0$. This concludes the proof.


\end{document}